\documentclass[a4paper,11pt]{article}
\usepackage{jcappub} 
\usepackage{lineno}
\usepackage{multirow}
\usepackage{multicol}
\usepackage{booktabs}
\usepackage{amsmath}
\usepackage{microtype}
\usepackage[table]{xcolor}
\usepackage{graphicx}

\title{\boldmath Classification of Spiral Galaxies by Spiral Arm Number using Convolutional Neural Network}

\author{Ming Wei Lee.}
\author{John Y. H. Soo}
\author{and Syarawi M. H. Sharoni}
\affiliation{School of Physics, Universiti Sains Malaysia, 11800 USM, Pulau Pinang, Malaysia}

\emailAdd{johnsooyh@usm.my}

\abstract{The structural information of spiral galaxies such as the spiral arm number, offer valuable insights into the formation processes of spirals and their physical roles in galaxy evolution. We developed classifiers based on convolutional neural networks (CNNs) using variants of the EfficientNet architecture with different transfer learning techniques and pre-trained weights to categorise spiral galaxies by their number of spiral arms. A selected dataset from Galaxy Zoo 2, comprising 11\,718 images filtered based on appropriate criteria is used for training and evaluation. Both the V2M model (EfficientNetV2M architecture fine-tuned on ImageNet) and the B0 model (EfficientNetB0 architecture with Zoobot pre-trained weights) achieved high accuracy on the down-sampled dataset, with most performance metrics exceeding 0.8 across all classes, except for galaxies with 4 arms due to the limited number of samples in this category. Merging higher-arm-number classes (more than 4 arms) improved the V2M model's accuracy significantly for 4-arm galaxies, as this approach allowed the model to focus on more distinct features within fewer, broader categories with a more balanced class distribution. GradCAM++ and SmoothGrad highlight the networks' effectiveness in classifying galaxies, through the distinction of the galaxy structures and the extraction of the spiral arms, with the V2M model showing better capabilities in both tasks. Lower-arm galaxies tend to be misclassified as "can't tell" when their spiral arms are not clearly visible, while higher-arm galaxies tend to be misclassified as having fewer arms when their features are only partially detected. The study also found that galaxies with 3 arms tend to have lower stellar masses, and this tendency is reduced in the model predictions. The models' mispredictions between 2-arm and 1/3-arm are likely resulting from external interference and dynamic nature of spiral arms. The V2M model prediction also shows a slight tendency towards higher stellar mass in high-arm galaxies.}

\begin{document}
\maketitle
\flushbottom

\section{Introduction}\label{refsec1}
\subsection{Galaxy Morphology}
Galaxy morphology is a description of galaxy structure that provides the understanding of galaxy formation, evolution and interaction with the environment. The Hubble sequence \citep{hubble_extragalactic_1926} is a fundamental classification scheme that mainly classifies a galaxy as an elliptical or a spiral. Spiral galaxies are subdivided according to bulge prominence and spiral arm tightness but researches suggested that there is weak correlation between the two \citep{kennicutt_star_1998,masters_galaxy_2019}. The de Vaucouleurs classification scheme \citep{devaucouleurs_classification_1959} refined spiral galaxies by including diffuse, irregular spiral arms and rings; while the Elmegreen classification \citep{elmegreen_flocculent_1982,elmegreen_arm_1987} categorised spiral galaxies into grand design (two well-defined spiral arms), multiple-armed (more than two well-defined spiral arms) and flocculent (many short and less-well defined arms). 

\subsection{Spiral Galaxy Formation}
According to the Galaxy Zoo and Galaxy Zoo 2 projects (GZ1 and GZ2 hereafter), spiral galaxies make up about two thirds of all massive galaxies \citep{lintott_galaxy_2011,willett_galaxy_2013}. Despite the prevalent existence of spiral galaxies in the local Universe, the exact mechanisms that initiate the formation of spiral arms are still not fully understood. The three main hypotheses are: (i) the quasi-stationary density wave theory \citep{lindblad_on_1953,lin_on_1964}, (ii) local instabilities that are swing amplified into spiral arms \citep{goldreich_spiral_1965,julian_non_1966}, and (iii) tidal interactions \citep{toomre_galactic_1972}. Bars may also play a role in inducing spiral arms, and these mechanisms are not necessarily mutually exclusive.

The density wave theory proposed that global spiral arms are the gradually developing patterns that rotate with fixed pattern speeds in the disc. Spiral arms arise from the propagation of waves with higher density across the disc, causing compression of gas as it traverses the wave, triggering star formation. However, the wave pattern will propagate outwards due to the differential rotational velocity and it could be maintained through swing amplification \citep{goldreich_spiral_1965}. In the swing amplification theory, the rotation of a density pattern from leading to trailing leads to the formation of spiral arms through a process of gravitational instability and amplification due to the shear induced by the differential rotation of the galactic disc. These spiral structures are transient, may dynamically exhibit trailing behaviour, and recurrently appear and disappear. This is supported by the tendency of trailing arm spirals observed in ref.~\citep{hubble_direction_1943}. The density wave theory primarily explains grand design spirals as their arms are relatively stable and long-lived, while swing amplification was proposed to elucidate multi-armed and flocculent structures. Simulations \citep{elmegreen_grand_1993,zhang_secular_1996} and observations \citep{block_morphological_1991,block_images_1994} have shown transient behaviours of spiral arms in isolated grand design galaxies. However, recent studies showed that the observed age gradient across a spiral arm is consistent with the prediction of the density wave theory \citep{shabani_search_2018,peterken_direct_2019,bialopetravi_study_2020,abdeen_evidence_2022}.

Tidal interactions occur due to the gravitational forces between two or more galaxies, often during galaxy mergers or near misses. Tidal arms are expected to have a pattern that winds up slower than transient arms, but are shorter-lived than quasi-stationary spiral arms. Observations found that the majority of grand design galaxies had companions or bars \citep{kormendy_observational_1979,elmegreen_flocculent_1982}, a result that is also commonly observed in simulations. One can conclude that grand design spirals are driven by density wave theory or tidal interaction.

While significant advancements have been made in understanding spiral galaxies, much of the focus has centred on two-armed systems, as around 60\% of observed galaxies exhibit some grand design structure \citep{elmegreen_flocculent_1982,grosbol_spiral_2004}. This limits insights into the mechanisms driving the development and stability of the more complex higher-arm spiral structures as studied in ref.~\citep{grosbol_galactic_2018,khrapov_modeling_2021}, despite the fact that multi-arm spirals are the common outputs from simulation. High-arm spirals exhibit unique patterns that challenge conventional density wave-driven formation theories, highlighting complex mechanisms that remain less understood, such as dynamic resonances in swing amplification and non-linear gravitational interactions. Well-classified high-arm galaxies offer valuable opportunities to address these gaps.

The previously mentioned observational studies have been restricted to relatively small samples of galaxies due to the constraint of visual inspection. Recent studies explore the spiral mechanisms by correlating between physical properties of galaxies and spiral arm number using classification data from large galaxy catalogues such as the GZ projects. Ref.~\citep{hart_galaxy_2018} discovered that the spiral arm number of 40\% of unbarred spiral galaxies from GZ2 and the Spitzer Survey of Stellar Structure in Galaxies (S4G) can be predicted by swing amplification in relation to the masses and sizes of haloes, bulges, and discs, with the inclusion of a dark matter profile with some level of expansion. By comparing the distribution of the environment, stellar mass and colour with respect to spiral arm number from GZ2 data, ref.~\citep{hart_galaxy_2016} found that the most massive galaxies favour many-arm spiral structure (their discs have not been perturbed to induce two-arm spirals), a high fraction of two-arm spiral galaxies was observed in the highest density environments (tidal interaction), and many-arm galaxies are much bluer than two-arm galaxies (short-lived phase). Thus, a unified explanation accounting for all observed types of spiral structure remains elusive.

\subsection{Galaxy Morphological Classifier}
Galaxy morphology classification was initially carried out through visual inspection by human experts \citep{vandenbergh_new_1976,hubble_extragalactic_1926}. Automated classification correlates morphological parameters with galaxy types \citep{abraham_morphologies_1994,abraham_galaxy_1996,abraham_morphologies_1996,conselice_asymmetry_2000} while machine learning exploits input parameters to classify galaxies \citep{naim_automated_1995,lahav_neural_1996,owens_oblique_1996,bazell_ensembles_2001,goderya_morphological_2002,ball_galaxy_2004,huertascompany_robust_2008,huertascompany_robust_2009,huertascompany_revisitng_2011}. Ref.~\citep{banerji_galaxy_2010} implemented artificial neural networks (ANNs) to classify Sloan Digital Sky Survey \citep[SDSS,][] {york_sloan_2000} data into ellipticals, spirals and point sources using the GZ1 catalogue. Recently, galaxy classification using deep learning algorithms has gained more attention, and in particular, convolutional neural networks (CNNs) are governing most of the image classification tasks as it can directly utilise pixel input. CNN outperforms other machine learning methods for classifying GZ1 samples into ellipticals and spirals in ref.~\citep{cheng_optimizing_2020}. Promising classification results were achieved with the implementation of CNNs as shown in ref.~\citep{dieleman_rotation_2015,dominguezsanchez_improving_2018,zhu_galaxy_2019,kalvankar_galaxy_2020,cavanagh_morphological_2021}.

The different types of spiral arms hold the key to the theory of spiral structure formation, but the morphological classification of spiral galaxies has not been adequately explored. Ref.~\citep{hart_galaxy_2017} utilised the machine learning algorithm \textsc{sparcfire} \citep{davis_sparcfire_2014} while ref.~\citep{bekki_quantifying_2021} developed deep learning architecture U-Net \citep{ronneberger_unet_2015} for detecting spiral arms. These approaches provide insight into the structural intricacies of spirals, however they are also relatively sophisticated, computationally expensive, and require high-resolution data. Instead, simpler classifiers may facilitate broader research endeavours that require basic morphological classification without the necessity for detailed structure analysis, and they can be readily scaled to handle the unprecedented size of future large datasets such as Euclid \citep{laureijs_euclid_2011}, Nancy Grace Roman Space Telescope \citep{spergel_wide_2015} and Vera C. Rubin Observatory \citep[LSST,][]{ivezic_lsst_2019}. Ref.~\citep{dieleman_rotation_2015} could not achieve a satisfactory classification accuracy for the question of spiral arm number (refer to Q11 in Table~\ref{reftable1}) attributed to the dataset selection that includes low confidence data. Therefore, we hope to develop a CNN-based classifier that can accurately classify spiral galaxies based on spiral arm number using a small and clean dataset, as human visual inspection is notably inefficient.

The aim of this research to (i) analyse the relationship between the spiral arm number and physical properties of galaxies, (ii) categorise grand design and multi-armed galaxies to explore their formation mechanisms, and (iii) classify high-arm galaxies to gain insights into their complex structural formation processes.


\section{Data: Galaxy Zoo 2}\label{refsec2}
Galaxy Zoo (GZ) is a citizen-science project which provided the classification of SDSS galaxies via the effort of the general public \citep{lintott_galaxy_2008}. At its current version, GZ2 is an extension of GZ1 which involved 16 millions classifications of nearly $300\,000$ galaxies drawn from the SDSS database \citep{willett_galaxy_2013}, including finer morphological features such as bars, bulges, spiral structures and more that can correlate with galaxy properties. The primary image dataset of GZ2 comprises of resolved galaxies derived from the SDSS North Galactic Cap with magnitude $r\leq17$, apparent radius $\texttt{petroR90\_r}>3$ and redshift $0.0005<z<0.25$. The dataset was supplemented with additional galaxies from Stripe-82 \citep{annis_sloan_2014} for which deeper, co-added imaging is available. In this paper, we used the main sample described in ref.~\citep{willett_galaxy_2013} which consists of $234\,500$ galaxies from SDSS Data Release 7 (DR7) and the Stripe-82 survey at normal depth imaging. The images of individual galaxies were cut out into $424\times424$ pixels, composed of \textit{gri} colours and scaled to $(0.02\times\texttt{petroR90\_r})$ arcsec per pixel for further processing.

The GZ2 decision tree\footnote{\url{https://data.galaxyzoo.org/gz_trees/gz_trees.html}} acts a hierarchical system guiding volunteers through a series of questions with multiple answers to classify galaxies based on their morphology, with each step leading to either another question or the conclusion of the classification process. The decision tree has 11 questions and 37 possible responses in total, as illustrated in Table~\ref{reftable1}. GZ2 provides three methods to aggregate the classifications for each task, they are vote fraction, weighted vote fraction and debiased vote fraction \citep{willett_galaxy_2013}. In this study, we make use of both the weighted $(\rho)$ and debiased ($\rho_m$) vote fraction values, where the former is computed by correcting the raw vote fractions with a function that down-weights classifiers in the tail of low consistency, while the latter adjusted to be independent of galaxy evolution due to redshift. We note that a different definition of the debiased vote fraction based on the new redshift-debiasing method provided in ref.~\citep{hart_galaxy_2016} is not considered in this study, as they provide extra information not readily available in the images. The ground truth for each task is selected as the answer with the highest weighted vote fraction in GZ2. We then focus on classifying spiral galaxies by the number of spiral arms, using question T11 from the GZ2 decision tree, which has 6 possible responses (see Table~\ref{reftable1}): 1 arm, 2 arms, 3 arms, 4 arms, more than 4 arms, and "can't tell". In this work, we use $m=1, 2, 3, 4, 5+$ to represent the spiral arm number, and use $m=?$ to represent a class of galaxies with unknown number of spiral arms ("can't tell").

\begin{table}[ht!]
\centering
\caption{Task questions of the GZ2 decision tree. The reader could refer to \url{https://data.galaxyzoo.org/gz_trees/gz_trees.html} for the full GZ2 decision tree.}
\label{reftable1}
\begin{tabular}{cp{5.8cm}cc}
\toprule
\textbf{Task} & \textbf{Question} & \textbf{Responses} & \textbf{Next} \\
\midrule
01 & & smooth & 07 \\
   & & features or disc & 02 \\
   & \multirow{-3}{5.8cm}{\textit{Is the galaxy simply smooth and rounded, with no sign of a disc?}} & star or artifact & end \\
\midrule
02 & & yes & 09 \\
   & \multirow{-2}{5.8cm}{\textit{Could this be a disc viewed edge-on?}} & no & 03 \\
\midrule
03 & & yes & 04 \\
   & \multirow{-2}{5.8cm}{\textit{Is there a sign of a bar feature through the centre of the galaxy?}} & no & 04 \\
\midrule
04 & & yes & 10 \\
   & \multirow{-2}{5.8cm}{\textit{Is there any sign of a spiral arm pattern?}} & no & 05 \\
\midrule
05 & & no bulge & 06 \\
   & & just noticeable & 06 \\
   & & obvious & 06 \\
   & \multirow{-4}{5.8cm}{\textit{How prominent is the central bulge, compared with the rest of the galaxy?}} & dominant & 06 \\
\midrule
06 & & yes & 08 \\
   & \multirow{-2}{5.8cm}{\textit{Is there anything odd?}} & no & end \\
\midrule
07 & & completely round & 06 \\
   & & in between & 06 \\
   & \multirow{-3}{5.8cm}{\textit{How rounded is it?}} & cigar-shaped & 06 \\
\midrule
08 & & ring & end \\
   & & lens or arc & end \\
   & & disturbed & end \\
   & & irregular & end \\
   & & other & end \\
   & & merger & end \\
   & \multirow{-7}{5.8cm}{\textit{Is the odd feature a ring, or is the galaxy disturbed or irregular?}} & dust lane & end \\
\midrule
09 & & rounded & 06 \\
   & & boxy & 06 \\
   & \multirow{-3}{5.8cm}{\textit{Does the galaxy have a bulge at its centre? If so, what shape?}} & no bulge & 06 \\
\midrule
10 & & tight & 11 \\
   & & medium & 11 \\
   & \multirow{-3}{5.8cm}{\textit{How tightly wound do the spiral arms appear?}} & loose & 11 \\
\midrule
11 & & 1 & 05 \\
   & & 2 & 05 \\
   & & 3 & 05 \\
   & & 4 & 05 \\
   & & more than four & 05 \\
   & \multirow{-6}{5.8cm}{\textit{How many spiral arms are there?}} & can't tell & 05 \\
\bottomrule
\end{tabular}
\end{table}

\subsection{Dataset Selection Criteria}
\begin{figure}[ht!]
\centering
\includegraphics[width=0.8\linewidth]{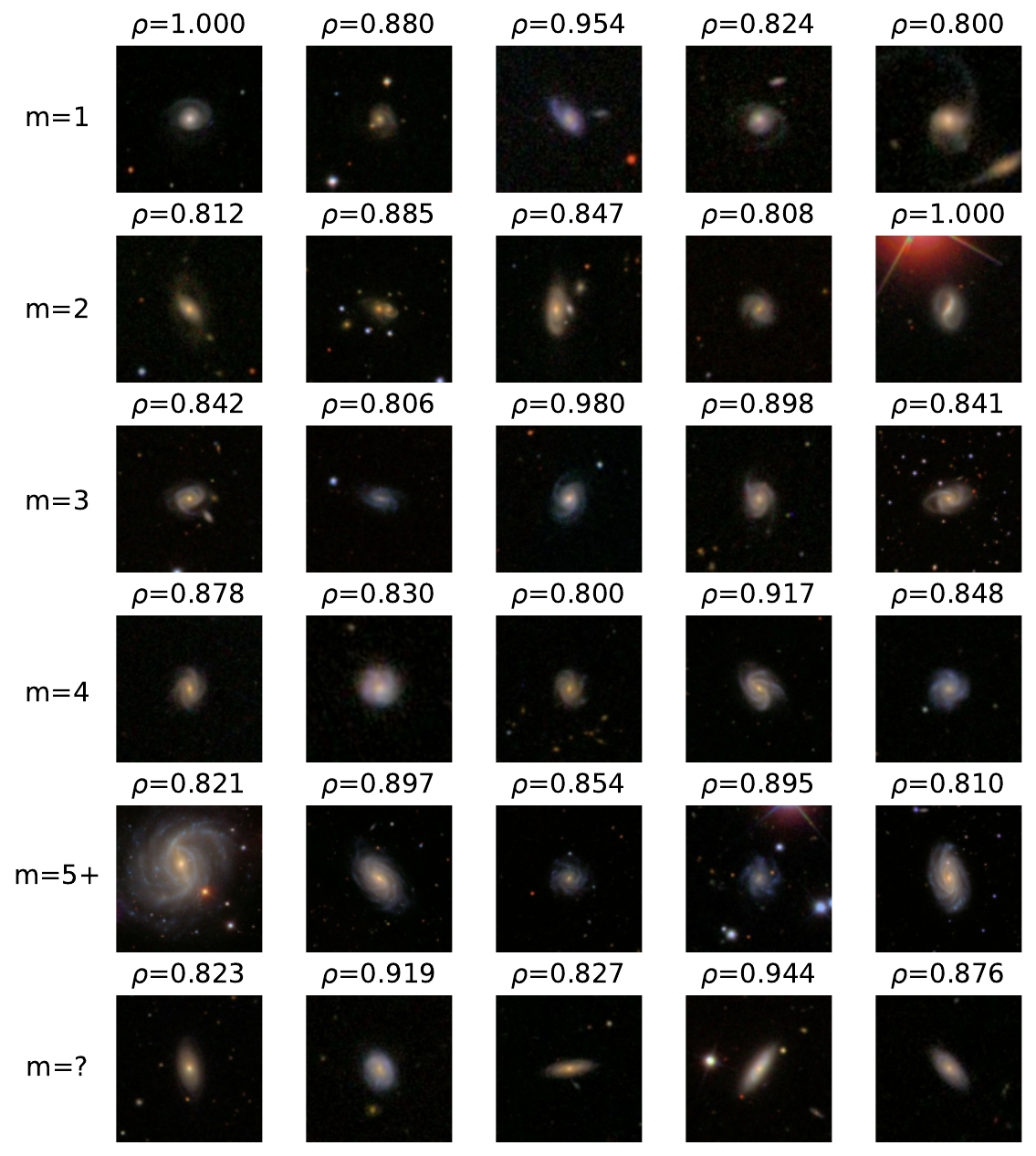}
\caption{Examples of galaxy images from our dataset with their respective debiased vote fraction, $\rho_m$. From top to bottom, each row represents the classes $m=1, 2, 3, 4, 5+, ?$.}
\label{reffigure1}
\end{figure}

In this study, we considered only galaxies from the clean and debiased sample with clear morphological classification and minimal false positives as explained in ref.~\citep{willett_galaxy_2013}. These pure objects can be selected such that they match a specific morphological category with appropriate criterion from the GZ2 main sample as shown in ref.~\citep{willett_galaxy_2013}. First, the weighted vote fraction for the preceding responses $(\rho)$ needs to achieve a certain threshold that is considered conservative to select clean objects. Next, the number of votes for the targeted question $(N)$ has to be greater than a certain number to be considered sufficient. Lastly, the debiased vote fraction for the targeted response ($\rho_\text{m}$) must achieve a baseline of 0.8. In this work, we select spiral galaxies with any number of spiral arms using the combination of cuts of $\rho_\textrm{features/disc}>0.430$, $\rho_\textrm{not edge-on}>0.715$, $\rho_\textrm{spiral,yes}>0.715$, $N_\textrm{spiral,yes}>20$ and $\rho_\text{m}>0.8$. The sample size reduces to 11\,718 images after filtering, and the classes $m=1,2,3,4,5+,?$ each contain a subsample of 165, 10677, 298, 28, 31 and 519 galaxies, respectively. Figure~\ref{reffigure1} shows sample images of each class with their respective $\rho_m$.

\subsection{Pre-processing}
\begin{figure}[ht!]
\centering
\includegraphics[width=0.75\linewidth]{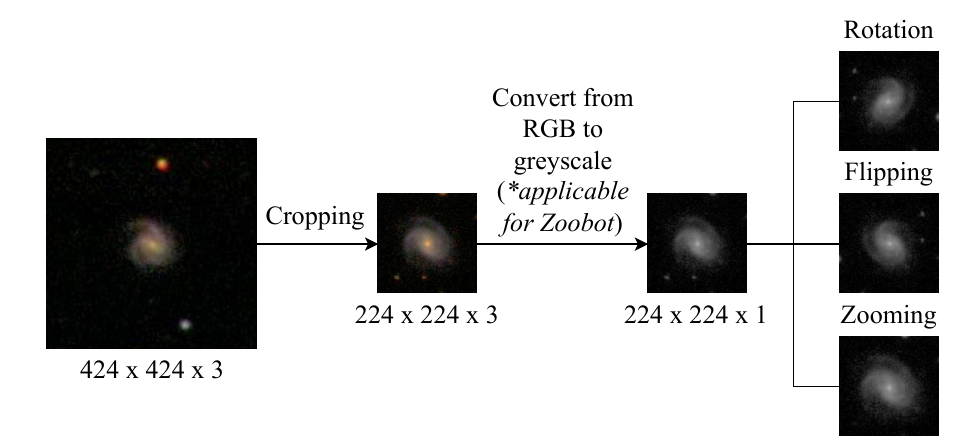}
\caption{Image pre-processing flow. The last three processes (rotation, flipping and zooming) are included for data augmentation.}
\label{reffigure2}
\end{figure}

The images obtained from GZ2 show a large field view through the telescope with the galaxy of interest at the centre. It is useful to crop the images to remove the large amount of sky background to avoid wrong feature extraction due to noises or secondary objects, this also reduces the dimensionality of the input and speed up model training. As shown in Figure~\ref{reffigure2}, images are cropped from $424\times424$ pixels to $224\times224$, with respect to the image centre. This cropping size is chosen based on the optimisation of model performance we conducted that will be further explained in Section~\ref{refsubsec4,1}. The images are converted from 3 RGB channels into 1 greyscale channel for the Zoobot model (explained in Section~\ref{refsec3}). It has been established in our preliminary experiments that training on data with coloured input images do not exert a notable impact on the model performance. The input images are linearly normalised such that pixel values are between 0 and 1 instead of 0 and 255 before passing into the Zoobot model as the pre-trained model receives pixel values in that range.

\subsection{Data Augmentation}
In order to avoid overfitting when training a model with a limited dataset, we resorted to the effective method of data augmentation. A model is overfitted when it performs too well on training data but fails to generalise in unseen data. Data augmentation is used to artificially increase the size of a training dataset by applying transformations and introducing variability to the existing training data. Rotation invariance of galaxy images was achieved in ref.~\citep{dieleman_rotation_2015} by exploiting translational and rotational symmetry in the CNN model. In this paper, three different forms of data augmentation were used to enforce a degree of spatial and translational invariance, as shown in Figure~\ref{reffigure2}. First, random rotation is applied to the training images in the range of \([-180^{\circ},180^{\circ}]\). Next, training images are flipped randomly. Lastly, training images are zoomed in or out with a factor in the range of 20\% of the original image size. Pre-processing will not be conducted at the testing phase.

\section{Algorithm}\label{refsec3}
\subsection{Convolutional Neural Network (CNN)}
\begin{figure}[t!]
\centering
\includegraphics[width=0.65\linewidth]{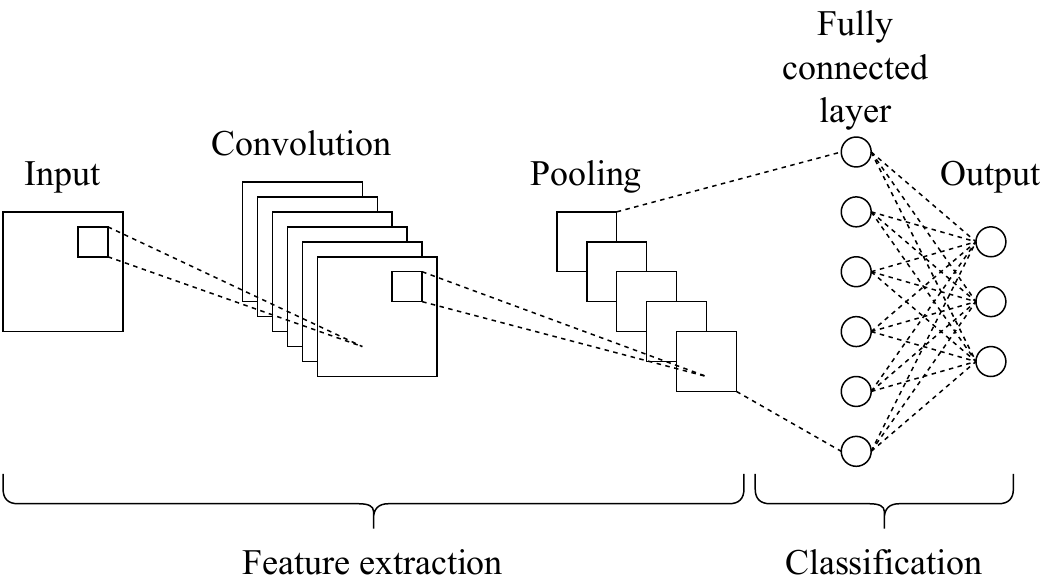}
\caption{Schematic diagram of a basic convolutional neural network architecture.}
\label{reffigure3}
\end{figure}

The CNN \citep{lecun_handwritten_1989} is a class of deep neural networks which takes multi-array data, particularly pixels in images as inputs. A CNN has a similar architecture as an artificial neural network (ANN), but the hidden layers are built up by convolutional layers, pooling layers and fully connected blocks \citep{goodfellow_deep_2016} as shown in Figure~\ref{reffigure3}. A convolutional layer applies a set of learnable filters to the input data and produces an output feature map. Each filter is a matrix of numbers that slides over the input data; discrete convolution can be seen as a dot product between the filters and input data. A convolutional layer can be computed as follows,
\begin{equation}
\label{refeq1}
x_{j}^{\ell} = f \left( \sum_{i \in M^{\ell}} x_{i}^{\ell} k_{ij}^{\ell} + b_{j}^{\ell} \right)    
\end{equation}
where $\ell$ is the layer number, $f$ the activation function, $k$ the convolutional kernel, $M$ the receptive field and $b$ the bias. The receptive field is a restricted sub-area for a neuron to receive input from the previous layer. A pooling layer reduces the dimensionality of a feature map by computing some aggregation function across small local regions of the input. A fully-connected layer connects every neuron in the previous layer to every neuron in the next layer and performs a linear transformation followed by an activation function. For more info on the full workings of a CNN, the reader is referred to ref.~\citep{lecun_gradient_1998,gu_recent_2018}.

In this work, we train two models, both based on the ready-made EfficientNet architectures, which outperformed other deep learning architectures in our preliminary experiments. One model is fine-tuned from the EfficientNetV2M model pre-trained on ImageNet, and another feature extracted from the Zoobot model with an EfficientNetB0 backbone pre-trained on galaxy images. They are explained in the following sections.

\subsection{EfficientNet}
Residual networks were introduced by ref.~\citep{he_deep_2016} to overcome the problem of vanishing gradients and degradation of accuracy in deep CNNs by using skip connections. Skip connections allow the network to learn residual functions with reference to the layer inputs, instead of learning from the underlying mappings of a few layers. EfficientNet is a CNN architecture and scaling method that uniformly scales all dimensions of depth, width and resolution in a principled way using a set of fixed scaling coefficients \citep{tan_efficientnet_2019}. It is proposed to address the issue of computational efficiency and gradient vanishing when training deep or wide CNNs after the introduction of the Residual Neural Network (ResNet). The architecture of EfficientNet is based on the inverted bottleneck residual blocks of MobileNetV2 \citep{sandler_mobilenetv2_2018}, in addition to squeeze-and-excitation blocks that explicitly modelled the dependencies between channels of its convolutional filters \citep{hu_squeeze_2018}. Ref.~\citep{tan_efficientnetv2_2021} improve the original EfficientNet by (i) using smaller image sizes for training to reduce the memory usage, (ii) replacing mobile inverted bottleneck convolution (MBConv) with fused inverted bottleneck convolution (Fused-MBConv) to utilise server accelerator, (iii) implementing training-aware neural architecture search (NAS) to search for best combination of MBConv and Fused-MBConv, and (iv) using modified progressive learning to solve the drop of accuracy.

EfficientNetB0 to B7 models are introduced in ref.~\citep{tan_efficientnet_2019}, with B0 being the smallest model and B7 the largest. EfficientNetV2S, EfficientNetV2M and EfficientNetV2L are introduced later in ref.~\citep{tan_efficientnetv2_2021}, where V2S is the smallest and V2L is the largest model.

\subsection{Zoobot}
Deep learning models such as MobileNet \citep{sandler_mobilenetv2_2018}, ResNets \citep{he_deep_2016} and EfficientNet usually incorporate a pre-trained classification backbone that are trained on terrestrial image datasets such as ImageNet \citep{deng_imagenet_2009}. Transfer learning is a common technique in deep learning applications where the pre-trained model is used as the feature extractor to capture high-level representations of input images then transferred and fine-tuned for a specific downstream task. This allows the new model to achieve better performance with less data and computation compared to training from scratch.

In this study, we compare the generic classification backbone pre-trained on ImageNet with a domain-specific CNN named Zoobot \citep{walmsley_zoobot_2023}, which was pre-trained on galaxy images. Zoobot is a Python package that utilises deep learning models like EfficientNet, DenseNet \citep{huang_densely_2017} and ResNet to predict GZ decision tree responses. A variant of Zoobot based on the EfficientNetB0 architecture is employed in the current study, which has been trained to classify galaxies from the Dark Energy Camera Legacy Survey (DECaLS) imaging data based on GZ responses \citep{walmsley_galaxy_2022}. Ref.~\citep{walmsley_practical_2022} demonstrated that the model pre-trained on a complex classification task can be extended to classify ring galaxies based on a smaller dataset that are out of its original training purpose. Ref.~\citep{etsebeth_astronomaly_2024} utilised the pre-trained CNN model from ref.~\citep{walmsley_practical_2022} to develop a galaxy anomaly detector via active learning. It is proven that a CNN backbone trained on astrophysical images outperform those trained on terrestrial images in astronomical object classifiers \citep{walmsley_practical_2022} and detectors \citep{popp_transfer_2024}.

\subsection{Network Architecture}
An EfficientNetV2M architecture \citep{tan_efficientnetv2_2021} fine-tuned from ImageNet pre-trained weights (hereafter referred to as V2M), and an EfficientNetB0 architecture feature extracted from Zoobot pre-trained weights (hereafter referred to as B0), will be used in the current study. For the B0 model, the pre-trained Zoobot model is used as a fixed feature extractor, where the learned weights of the model are frozen and not updated during training. Whereas for the V2M model, the pre-trained model will be fine-tuned by updating the weights of the layers to adapt to the new task. The differences in the hyper-parameters for the B0 and V2M models are listed in Table~\ref{reftable2}. Although the V2M model has deeper layers, the B0 model is expected to perform comparably since it is pre-trained on a galaxy-domain dataset.

\begin{table}[ht!]
\centering
\caption{Comparison of hyperparameters between the B0 and V2M models.}
\label{reftable2}
\begin{tabular}{lll}
\toprule
& B0 model & V2M model \\
\midrule
CNN backbone & EfficientNetB0 & EfficientNetV2M\\ 
\midrule
Number of hyparameters& $5\,366\,883$ & $54\,468\,282$\\ 
\midrule
Pre-trained weights& Zoobot & ImageNet\\
\midrule
Layers allowed for weight updates& Dense layers & All layers\\
\bottomrule
\end{tabular}
\end{table}

\begin{figure}[ht!]
\centering
\includegraphics[width=0.25\linewidth]{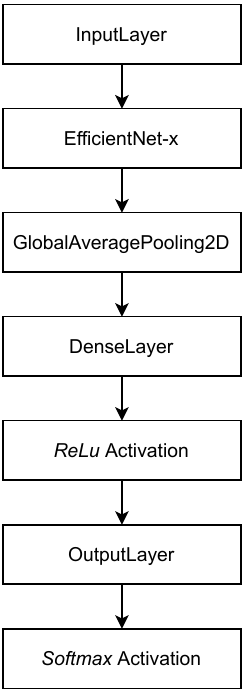}
\caption{Network architecture of the tail part of the classification models tested in this study. The variable 'x' here is a placeholder for both the B0 and V2M backbones.}
\label{reffigure4}
\end{figure}

A global average pooling layer that downsamples the output of every dimension and flatten the tensor is introduced to the tail part of EfficientNet architecture. The output of this non-linear transformation is passed onto a fully connected dense layer with 1028 connections via a ReLU activation \citep{agarap_deep_2018}. The output layer consists of 6 classes with a softmax activation function \citep{bridle_training_1989}, a type of normalised exponential which converts the activations into probabilities that sum to 1 across all output classes. The predicted output is the class with the highest probability. An overview of this network architecture is shown in Figure~\ref{reffigure4}.

\subsection{Implementation Details}
The network architectures are constructed with the high-level API Keras \citep{chollet_keras_2015} and trained using the Tensorflow library \citep{abadi_tensorflow_2016} running with Nvidia GPU programming toolkit CUDA \citep{nvidia_cuda_2020}. The programming codes are built using Python and training was carried out on Kaggle\footnote{\url{https://www.kaggle.com/code}} with 2 Nvidia T4 graphics cards. The dataset is split into training and validation sets with a ratio of 70:30. Training is conducted over a maximum of 50 epochs with a batch size of 16. Early stopping to cease training once there is no increment in the validation accuracy over 30 epochs is applied. This helps to avoid unnecessary computation cost once the model performance has plateaued. The best weights are stored using Keras's ModelCheckpoint callback function. Categorical cross-entropy loss function, Adam optimiser \citep{kingma_adam_2017} with a constant learning rate of 0.0001 and a Glorot uniform weight initialiser were used for model training.

\section{Performance Optimisation}\label{refsec4}
In this section, we introduce the performance metrics used in this work, and we show how our model's accuracy is enhanced by carefully selecting the optimal image cropping size and determining the appropriate number of images in the training dataset.
\subsection{Performance Metrics}
In most cases, accuracy is used to measure model performance during training, which can be formulated as the ratio of number of correctly classified data to the total number of data. However, accuracy alone is inadequate to measure the overall model performance especially when it comes to multi-class classification with an imbalanced dataset. A confusion matrix can be used to compare the ground truth label to the predicted labels. From here, the model performance can be further assessed by calculating precision, recall and F1-score, defined as below,
\begin{equation}
\text{Precision} = \frac{\text{TP}}{\text{TP} + \text{FP}}
\end{equation}

\begin{equation}
\text{Recall} = \frac{\text{TP}}{\text{TP} + \text{FN}}
\end{equation}

\begin{equation}
\text{F1} = 2 \times \frac{\text{Precision} \times \text{Recall}}{\text{Precision} + \text{Recall}}
\end{equation}
where TP, FP and FN denote the true positives, false positives and false negatives respectively. Precision is equivalent to the positive predictive value, recall is equivalent to true positive rate and F1-score is the harmonic mean of precision and recall. All of them have values between 0 and 1. In the subsequent sections, the classification results obtained from the validation set will be utilised for comprehensive analysis and discussion.

\subsection{Optimisation Results}
\subsubsection{Selection of Image Size}\label{refsubsec4,1}
In machine learning, it is crucial to have a trade-off between model performance and computational efficiency. One of the methods to achieve optimal configuration is by selecting the best image size for model training. The region of interest (ROI) in an image can capture different levels of spatial information and determine the feature representations learned by the model, hence affecting the model accuracy and generalisation ability. As shown earlier in Figure~\ref{reffigure1}, the galaxies occupy only the centre part of the image and most of them come along with secondary objects or noises. Thus, it is useful to remove the redundant background and avoid the noises to be learned as features during model training. 

A series of experiments were conducted in which binary classification models are trained on a subset of the total data to classify between $m=2$ and $m=3$ using various image sizes. In this context, the dataset is balanced between the 2 classes with a limit of 300 images per class. Due to the nature of inherent variability in deep learning models, the variability in model performance across multiple training runs by using the same training configuration is required to be accounted for. Five training runs are conducted for each image size and the standard deviation of the F1-score is calculated to quantify the variability of model performance. The coefficient of variation (CV) is used to statistically compare the relative variability with the average F1-score. We use the expression $\textrm{CV}=\sigma/\mu$, where $\sigma$ and $\mu$ denotes standard deviation and mean respectively.

\begin{figure}[t!]
\centering
\includegraphics[width=0.7\linewidth]{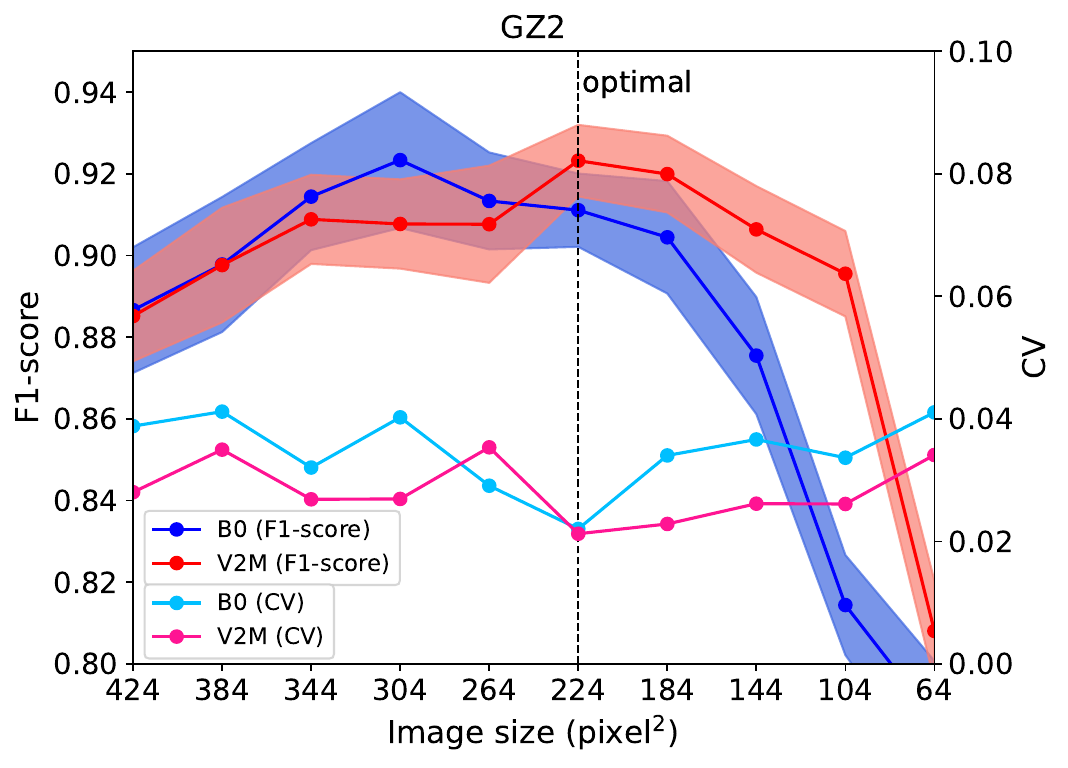}
\caption{Plot of F1-score (with its standard deviation) and CV across different image sizes for the B0 and V2M models. Both models achieve their optimal image size at $224\times224$.}
\label{reffigure5}
\end{figure}

Figure~\ref{reffigure5} shows the relationship between image size and F1-score/CV for the B0 and V2M models. The image size that consistently yields high F1-score across multiple training runs can be determined with the lowest CV value that suggests a favourable balance between stability and performance. As a result, the optimal image size is $224\times224$ for both the B0 and V2M models. The B0 model achieves an optimal F1-score of 0.9111 with a CV of 2.203, while the V2M model attains an optimal F1-score of 0.9232 with a CV of 2.124. The graphs of B0 and V2M models share the same trend where the F1-score increases gradually when the image size decreases and then dramatically drops after reaching an optimal. The B0 model attains its maximum point before the V2M model, but the instability of the F1-score, indicated by a high CV value, renders it unsuitable for selection as the optimal configuration. In general, the B0 model exhibits a wider range of CV value across image sizes compared to V2M model.

\subsubsection{Number of Images}\label{refsubsec4,2}
\begin{figure}[t!]
\centering
\includegraphics[width=0.75\linewidth]{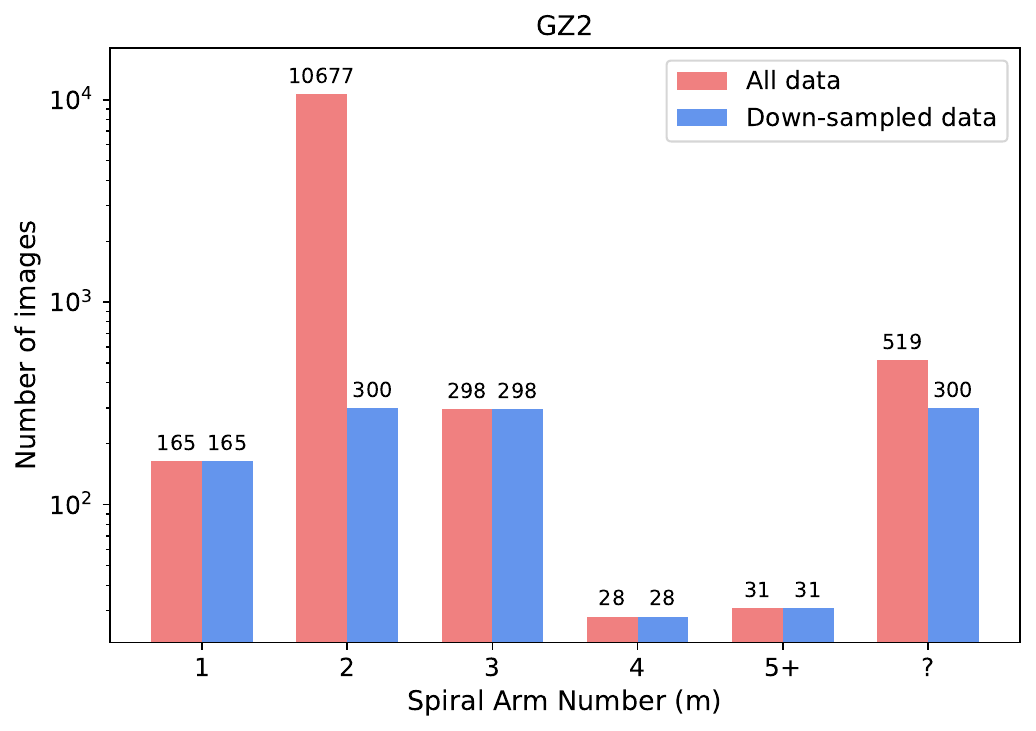}
\caption{Number of galaxy images for the full and down-sampled data sets according to their classes.}
\label{reffigure6}
\end{figure}

The distribution of images across different classes in our dataset is imbalanced, e.g. the class $m=2$ has over 10\,000 images, while $m=4$ and $m=5+$ each have less than 35 images. Two-armed galaxies represent the most prevalent category of spiral galaxies observed within our universe, thus resulting in a profusion of relevant images within the GZ2 dataset. The limited resolution of the astronomical survey utilised for the compilation of the GZ2 dataset, coupled with the considerable distances of celestial objects, presents a challenge to accurately capture the galaxies with more complex structures, particularly those of $m=4$ and above, which may appear fainter or more diffuse in the images. Down-sampling is a common technique used in addressing data imbalance by reducing the number of samples in the majority classes to create a more equitable distribution of data. An experiment was conducted to investigate the effect of data balancing on the performance of the B0 model, where the result of the model trained on all data is compared with the result of a down-sampled set of data which is limited to 300 images per class. In the process, the over-represented classes ($m=2$ and $m=?$) are randomly subsampled. Figure~\ref{reffigure6} shows the distribution of images for both the full and down-sampled data set used in this optimisation exercise.

The validation results of the B0 model trained on the full and down-sampled data are compared in Table~\ref{reftable3}. The precision of the model trained on all data exhibits high values across all classes, surpassing 0.8 overall. The $m=2,5+$ and $?$ classes each achieve a high F1-score above 0.8. Despite high precision values, both recall and F1-score are low for $m=1$, $3$ and $4$. When trained on imbalanced data, the class $m=4$ has the lowest F1-score (0.364), followed by $m=1$ (0.535) and $m=3$ (0.631). This discrepancy can be attributed to the heavily skewed distribution of the dataset towards $m=2$, resulting in misclassification of most samples to this class. 

\begin{table*}[ht!]
\centering
\caption{Precision, recall and F1-score for the B0 model trained on either all data (imbalanced) or down-sampled (balanced) data. The F1-score for class $m=1$ and $3$ using the down-sampled data improved by 63\% and 33\% respectively. The results for the V2M model trained on the down-sampled data is shown for comparison.}
\label{reftable3}
\resizebox{\textwidth}{!}{%
\begin{tabular}{ccccccccccc}
\toprule
& \multicolumn{7}{c}{B0} & \multicolumn{3}{c}{V2M} \\
\cmidrule(lr){2-8} \cmidrule(lr){9-11}
& \multicolumn{3}{c}{All Data} & \multicolumn{3}{c}{Down-sampled Data} & & \multicolumn{3}{c}{Down-sampled Data}\\
\cline{2-7}
\multirow{-3}{*}{m} & Precision & Recall & F1-score & Precision & Recall & F1-Score & \multirow{-2}{*}{Diff. in F1-Score} & Precision& Recall& F1-Score\\
\midrule
1 & 0.905 & 0.380 & 0.535 & 0.932 & 0.820 & 0.872 & 63.0\%&0.909&0.800&0.851\\
\midrule
2&0.969&0.989&0.979&0.811&0.856&0.832&-15.0\%&0.835&0.900&0.866\\
\midrule
3&0.797&0.522&0.631&0.835&0.844&0.840&33.1\%&0.862&0.900&0.880\\
\midrule
4&1.000&0.222&0.364&1.000&0.222&0.364&0.0\%&1.000&0.222&0.364\\
\midrule
5+&0.818&0.900&0.857&1.000&0.900&0.947&10.5\%&0.769&1.000&0.870\\
\midrule
?&0.831&0.821&0.826&0.878&0.956&0.915&10.8\%&0.944&0.933&0.939\\
\bottomrule
\end{tabular}%
}
\end{table*}

Conversely, all performance metrics of the model trained on the balanced down-sampled data show improvements across all classes, except for $m=2$ ($-$15\%). The class $m=1$ has the highest improvement (63\%), follows by $m=3$ (33\%) and $m=5+$, $?$ (10\% each). Nevertheless, the F1-score for $m=2$ remains above 0.8, indicative of satisfactory performance. In the case of $m=4$, there are no changes in the performance metrics. Overall, the down-sampled configuration offers the potential to develop a more robust classifier capable of accurately classifying spiral galaxies according to arm number, therefore we will use this configuration in the following sections.

\section{Results and Discussion} \label{refsec5}
\subsection{Model Comparison}
For our main results, the B0 and V2M models are trained on the down-sampled dataset mentioned in~\ref{refsubsec4,2} to classify galaxies into spiral arm numbers $m=1,2,3,4,5+$ and $?$, and the results were presented in Table~\ref{reftable3}. Overall, the V2M model exhibits similar results as compared to the B0 model, with all the metrics having values greater than 0.8 except for $m=4$ and $5+$. In the case of $m=5+$, the precision value is slightly below 0.8. The V2M model marginally outperforms the B0 model for $m=2,3$ and $?$, but also performs poorly in classifying $m=4$ galaxies, mirroring the performance of the B0 model. The recall and F1-scores are only slightly higher than random (0.222 and 0.364 respectively) for both models, despite the fact that the precision is considerably high. This is indicative of the challenge in accurately classifying classes with insufficient image samples.

Our achieved results surpass those of ref.~\citep{dieleman_rotation_2015}, showing over a 100\% improvement in F1-score for $m=1$, $3$ and $5+$, as well as an increase of over 40\% for $m=?$, despite differences in the data selection criteria. In their work, the samples were selected based on the cumulative probability derived from the debiased weighted vote fraction, which was adjusted by multiplying the debiased weighted vote fraction of the responses with the value which led to that question. This approach assigned greater weights to more fundamental morphological categories higher in the decision tree, and a threshold of 0.5 was used to select the samples. Generally speaking, our selection criteria is tighter, resulting in cleaner samples.

\subsection{Class Combination}
\begin{figure*}[t!]
\centering
\includegraphics[width=\linewidth]{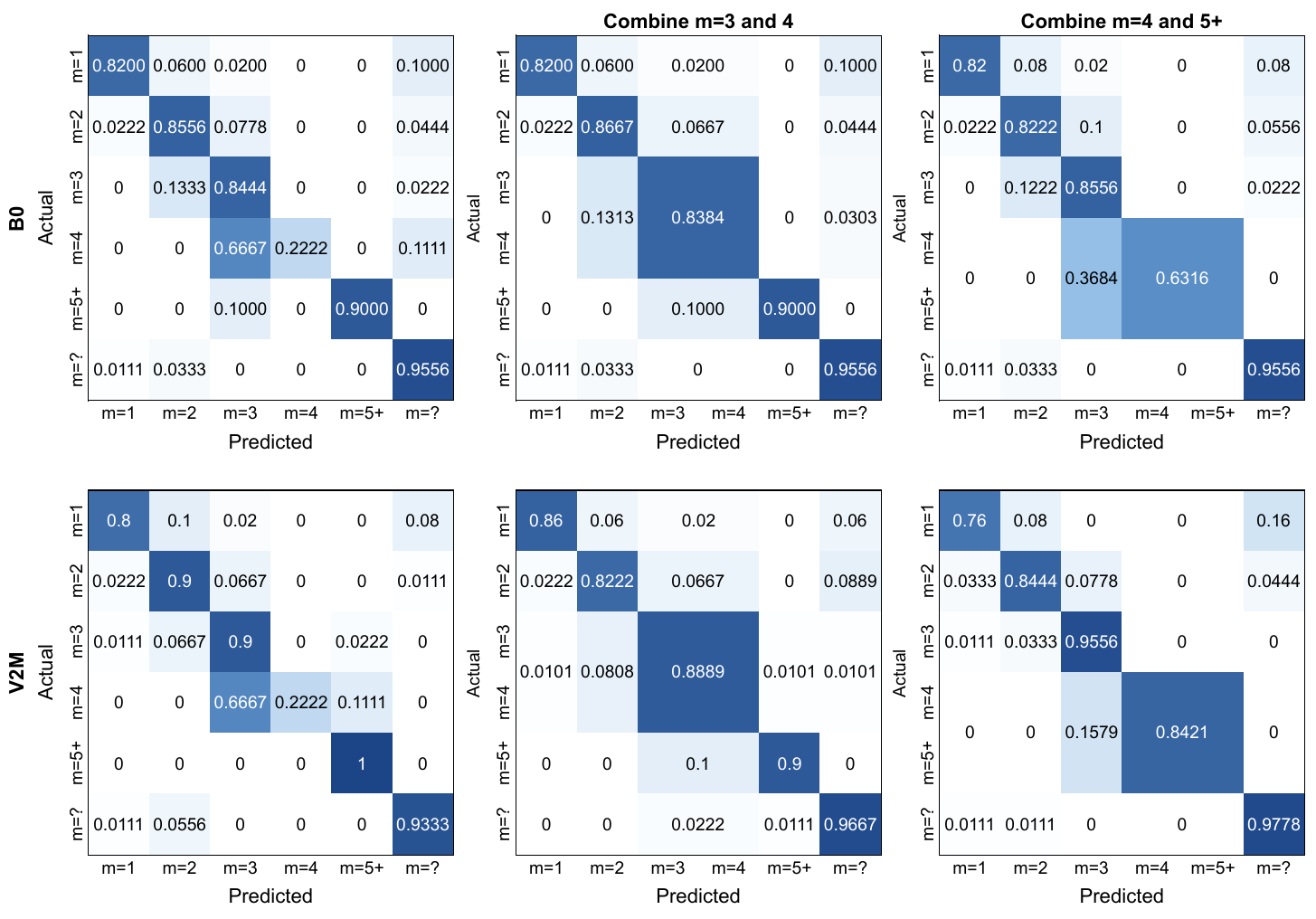}
\caption{The confusion matrices of the B0 model (top) and the V2M model (bottom) for 3 types of dataset configuration : all classes (left), combined  $m=3$ and $4$ (centre) and combined $m=4$ and $5+$ (right). Up to 22\% of $m=3$ galaxies are misclassified as $m=4$ by both the B0 and V2M models under the first configuration. The misclassification rate decreased to under 20\% when using the second and third configurations for the V2M model.}
\label{reffigure7}
\end{figure*}

In multi-class classification, it is essential to visualise results via the presentation of a confusion matrix, as this allows us to anticipate which distinctions are most problematic and likely to be sources of confusion. In an individual confusion matrix, each column represents the instances in a predicted class while each row represents the instances in the class labelled in GZ2. Values on the matrix diagonal indicate correct predictions while the values outside of that show incorrect prediction. The normalised confusion matrices of both the B0 and V2M models for the validation set are presented in the first column of Figure~\ref{reffigure7}. For those classes with results better than the baseline of 0.8, i.e, $m=1,2,3,5+$ and $?$, the misclassified galaxies are almost equally distributed among the classes with greater sample count, i.e. classes other than $m=4$ and $5+$. This is likely due to a combination of high-population categorical bias. A considerably high percentage of $m=3$ galaxies are misclassified as $m=4$ (67\%) for both B0 and V2M models while the remaining 11\% are misclassified as $m=?$ for B0 while $m=5+$ for V2M. 

It is observed that in a majority of instances, both the B0 and V2M models tend to misclassify galaxies with 4 spiral arms as those having 3 spiral arms. To improve the classification of these classes, various configurations are tested by combining different classes, as described in the second and third column of Figure~\ref{reffigure7}. We explored the possibility of combining  classes $m=3$ and $m=4$, and observed a significant improvement in performance for both the B0 and V2M models, with all classes achieving an accuracy of 0.8 or higher.

Alternatively, we tested a third configuration which combines $m=4$ and $m=5+$. We believe that this configuration better reflects the progression of galaxy structures and provides a more scientifically reasonable classification when compared to the second configuration. The $m=4$ and $5+$ galaxies exhibit structural similarities, particularly in their higher spiral arm numbers and more complex morphology, making them naturally aligned for classification. With this configuration, the B0 model does not show a significant improvement in the accuracy of the merged class, as 37\% of the combined class is still misclassified as $m=3$. In contrast, the V2M model produced much better results, with the accuracy of the combined class improving to over 0.8. 

While combining $m=3$ and $4$ could yield a higher improvement, the motivation for distinguishing between 3 and 4 arms remain scientifically significant, as these structure might be governed by different spiral arm formation mechanism. Therefore, we conclude that combining $m=3$ and $4$ offers the best results, however if a distinction between 3 and 4 arms is needed, the third configuration acts as a balance between scientific rigour and model performance, which is in sync with what we discussed in Section~\ref{refsec1}.

\subsection{Visualising the CNN}
\begin{figure*}[t!]
\centering
\includegraphics[width=0.7\linewidth]{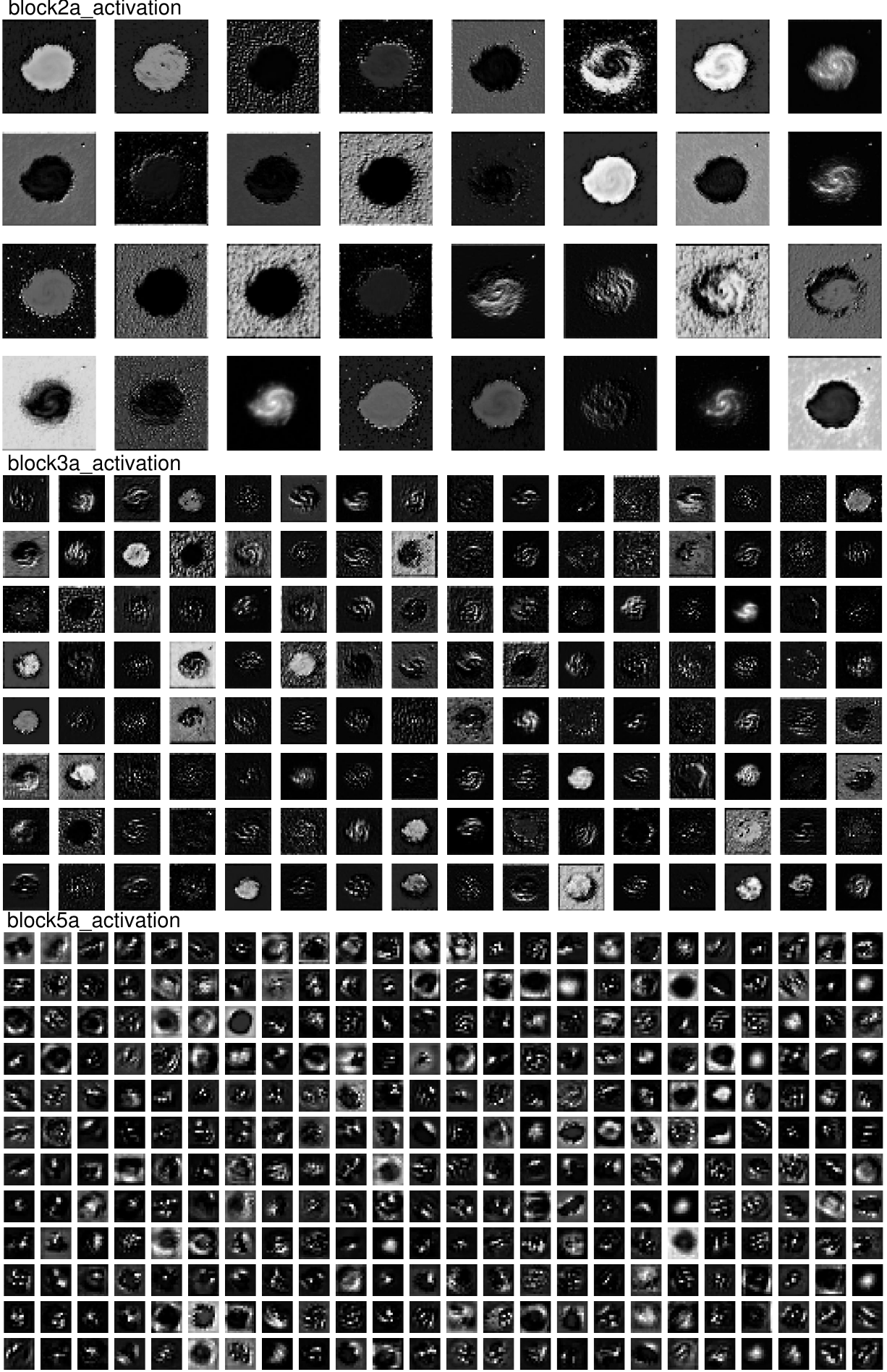}
\caption{Plot of feature map in the lower, middle and upper convolutional layers of the B0 model, as generated for a $m=2$ galaxy sample. The activations become more abstract in deeper network layers.}
\label{reffigure8}
\end{figure*}

To gain a deeper understanding of the networks' performance in classifying spiral galaxies, the activation maps in the immediate convolutional layers of the B0 model are visualised. These visualisations allow us to observe the progressive abstraction of features as the input image passes through the network. As an example, Figure~\ref{reffigure8} show the activation maps of a $m=2$ galaxy from the lower, middle and upper convolutional layers of the B0 model. Initially, the filters detect basic features such as edges and corners. As the network's depth increases, the filters capture more complex structures such as spiral arms and the central bulge of galaxies, and the activations become more abstract, combining basic features into higher-level representations.

To further elucidate the decision-making process of the CNN networks, Grad-CAM++ \citep{chattopadhay_gradcam_2018} and SmoothGrad \citep{smilkov_smoothgrad_2017} were employed to produce heatmaps that highlight the most relevant regions of an image to the network's prediction. Grad-CAM++ generates heatmaps using the gradients of class scores relative to the last convolutional layer and performs weighted pooling to highlight key image regions for class prediction, offering greater interpretability in higher-level features of galaxy morphology as illustrated in ref.~\citep{gordon_uncovering_2024,medinarosales_mitigating_2024}. On the other hand, SmoothGrad uses the saliency mapping approach to compute gradients of class scores with respect to the input image and measures the impact of small changes in each pixel on the class scores, offering pixel-level information as illustrated in ref.~\citep{bhambra_explaining_2022}.

\begin{figure*}[t!]
\centering
\includegraphics[width=\linewidth]{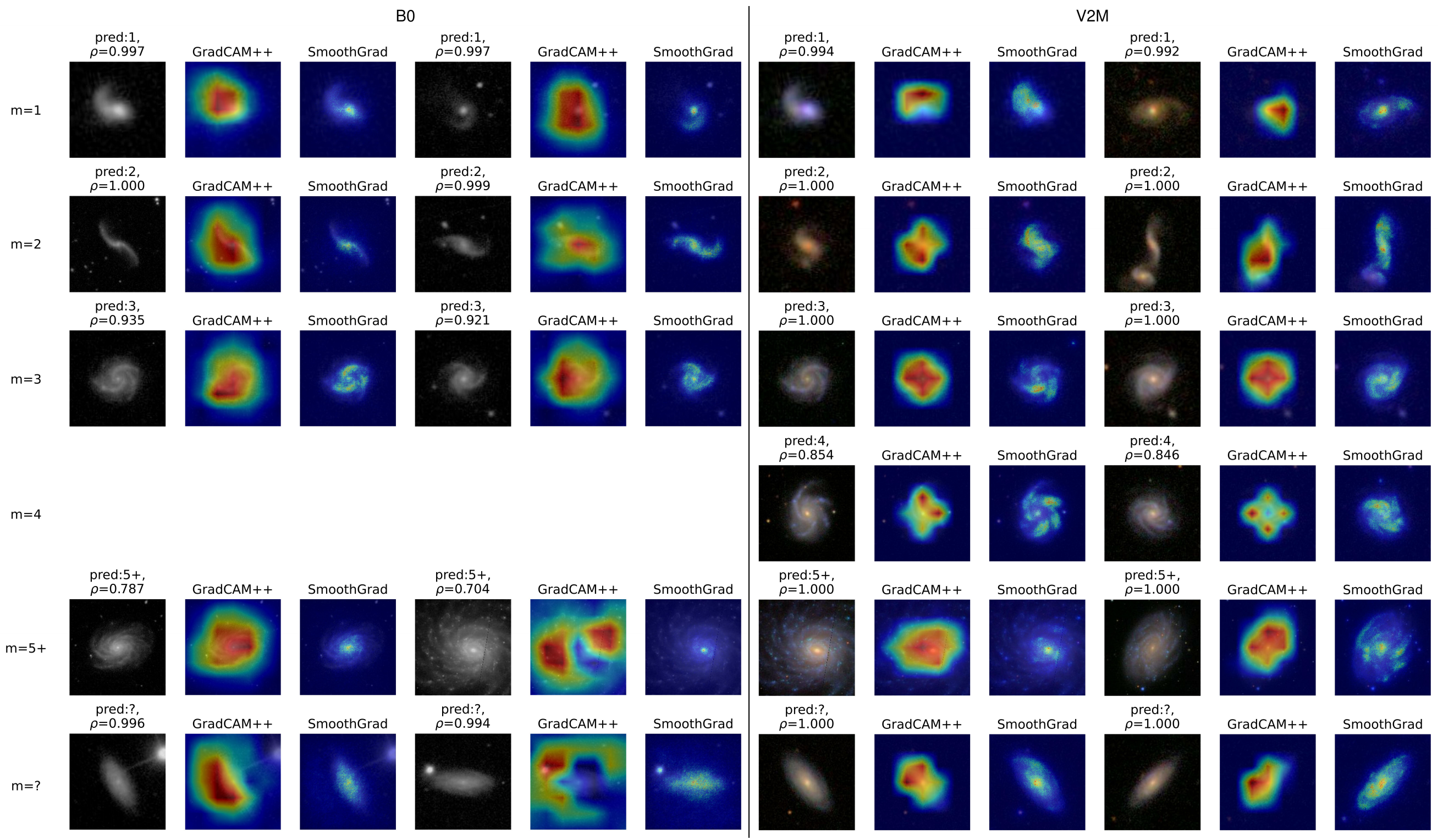}
\caption{Top-scoring galaxy samples with respective GradCam++ and SmoothGrad maps for each class, correctly classified by the B0 (left) and V2M (right) models, annotated with their classification probabilities. Empty rows indicate insufficient output of model training for the generation of heatmaps. A significant distinction is observed in $m=?$, where the B0 model highlights the background region rather than the galaxy itself.}
\label{reffigure9}
\end{figure*}

\begin{figure*}[t!]
\centering
\includegraphics[width=\linewidth]{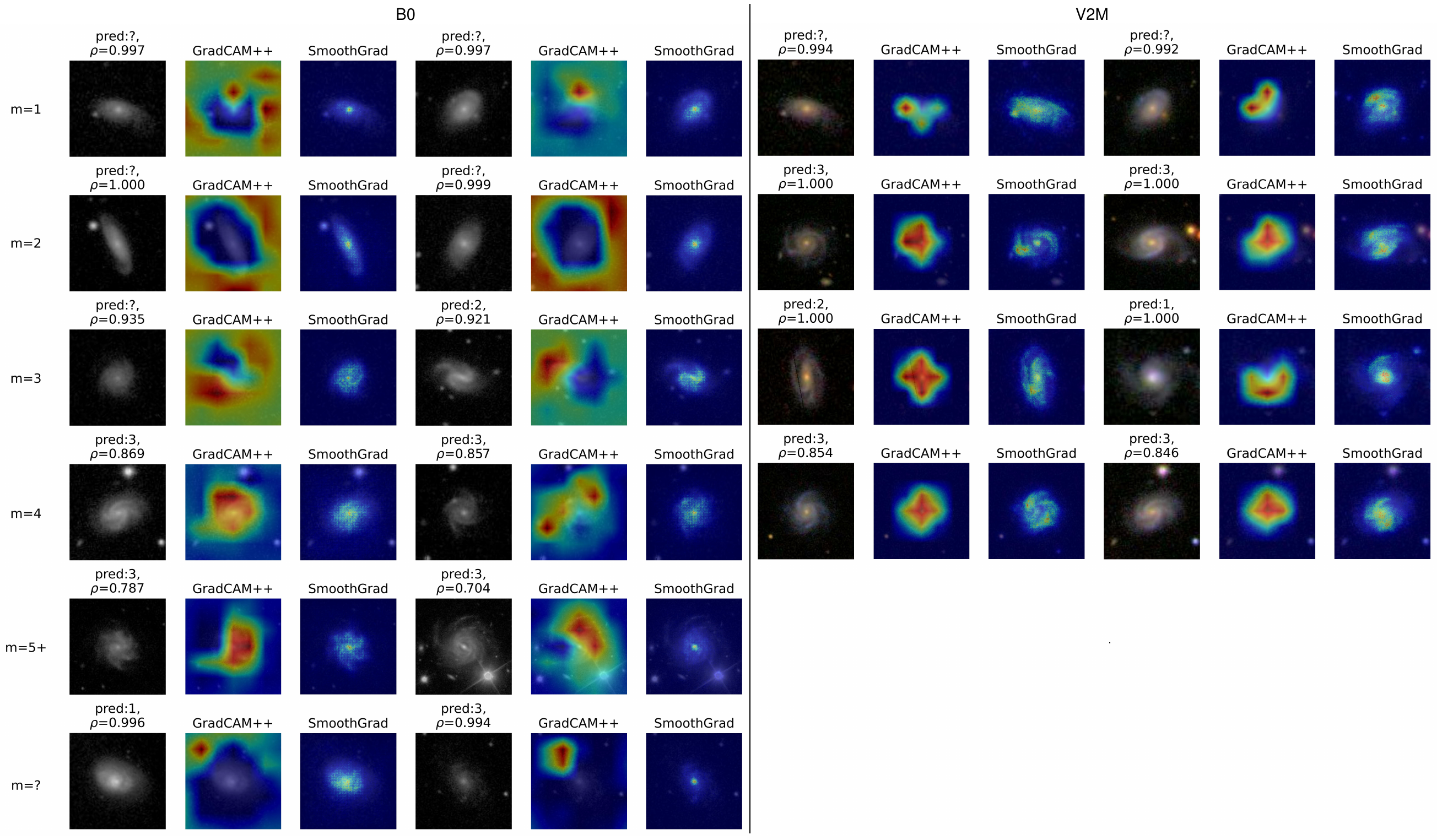}
\caption{Bottom-scoring galaxy samples with respective GradCam++ and SmoothGrad maps for each class, misclassified by the B0 (left) and V2M (right) models, annotated with their classification probabilities. See Figure~\ref{reffigure9} for more info.}
\label{reffigure10}
\end{figure*}

Figure~\ref{reffigure9} shows the top-scoring galaxies from our validation dataset with their corresponding Grad-CAM++ and SmoothGrad outputs, accurately predicted by the B0 and V2M models with the highest prediction probability. Figure~\ref{reffigure10}, on the other hand, displays galaxies from our validation dataset that were misclassified by the B0 and V2M models with the highest prediction probability. Empty rows indicate insufficient output from the model training for the generation of heatmaps. In the heatmaps, red regions indicate areas of high importance, yellow and green regions indicate moderate importance, and blue regions indicate little to no importance.


In Figure~\ref{reffigure9}, Grad-CAM++ highlights the entire galaxy structure for $m=1,2,3,4$ and $5+$ in the B0 model, suggesting that the model's decision making process is driven by the overall structural features of galaxies. For $m=?$, the model focuses on the surrounding region rather than the galaxy itself, implying that its classification is based on the absence of significant galaxy structural features. The Grad-CAM++ results of the V2M model highlight the galaxy features for all classes including $m=?$, indicating that its decision making process is also based on the differences in galaxy structures. Figure~\ref{reffigure10} demonstrates that low-arm galaxies are predicted as $m=?$ by the B0 model, where Grad-CAM++ highlights the surrounding region rather than the galaxy itself. This misclassification of galaxies as $m=?$ also happen to the V2M model when the spiral arms become diffuse and less visible. Grad-CAM++ results of the V2M model exhibit similar cross-shaped patterns in classes $m=2$,$3$ and $4$, leading to potential misclassification among these classes when the shapes are not sufficiently distinct for accurate differentiation. Higher-arm galaxies are misclassified as lower-arm galaxies when the galaxy structure is not completely captured by the models.

The SmoothGrad saliency maps in Figure~\ref{reffigure9} show that both models can extract spiral arms features, though this capability decreases for galaxies having many-arm and less visible arm structures. The V2M model has stronger ability to extract detailed structural information in spiral galaxies than the B0 model. The highlighted region in SmoothGrad results show some patterns across classes: a central bulge with extended arm in $m=1$, connected arms through the central bulge in $m=2$, detached multi-arms in $m=3$ and $4$, broken arm segments in $m=5+$, and a central bulge with diffuse arms in $m=?$. The misclassifications shown in Figure~\ref{reffigure10} occur when the spiral structure is misinterpreted as the pattern from other classes, demonstrating that accurate extraction of spiral arms results in correct classification, with incomplete extraction leading to misclassification. Galaxies with less visible arms tend to be predicted as $m=?$, as the models struggle to extract clear spiral arms. The class $m=5+$ tend to be misclassified as lower-arm galaxies, as the complexity of their structure cannot be fully captured by the models. The model's ability to learn complex spiral structures is likely limited by the small dataset of galaxies with class $m=4$ and $5+$. The extraction of spiral arms using SmoothGrad does not consider whether a spiral galaxy has well-defined or flocculent arms.

In summary, the V2M model is shown to demonstrate better capabilities in distinguishing galaxy structures and extracting detailed spiral features. This is likely due to the V2M model's deeper and enhanced architecture, which makes it more capable to learn intricate structures of spiral arms. Further fine-tuning could enable the model to adjust its pre-trained weights to better capture the complex spiral features. Conversely, using feature extraction with a frozen EfficientNetB0 backbone in the B0 model may be insufficient for classifying galaxies by spiral arm number, likely because the intricate features required were not sufficiently captured in the pre-trained Zoobot dataset, despite its higher similarity to the target dataset.

\section{Application: Spiral Arm Number vs. Stellar Mass} \label{refsec6}
Galaxy stellar mass has been shown to correlate with galaxy morphology \citep{bamford_galaxy_2009,kelvin_galaxy_2014}, as well as the Hubble type \citep{munozmateos_spitzer_2015}. Various works have demonstrated that the central mass in spiral galaxies influences the type of spiral structure they exhibit. Specifically, the total stellar mass in galaxies is linked to the observed spiral structures, where the strength of the class $m=2$ is more pronounced in galaxies with higher stellar mass \citep{kendall_spiral_2015}. Ref.~\citep{hart_galaxy_2016} correlated the number of spiral arms in GZ2 galaxies, classified using their new defined debiased vote fraction, with the stellar mass data from ref.~\citep{chang_stellar_2015}. They suggested that galaxies with higher stellar masses tend to have more spiral arms. Similarly, ref.~\citep{portertemple_galaxy_2022} conducted a study using the Galaxy And Mass Assembly - Kilo Degree Survey (GAMA-KiDS) DR3 dataset \citep{dejong_third_2017} and Multi-wavelength Analysis of Galaxy Physical Properties (MAGPHYS) stellar mass data \citep{dacunha_simple_2008} and suggested that galaxies with higher stellar mass tend to have more well-defined spiral arms. 

This section explores the correlation between galaxy stellar mass and spiral arm number, using the data from GZ2 labels and model predictions, with $m=4$ and $5+$ combined. We utilise images from our entire data set (including all training, validating and testing images), giving a total of 11\,718 images. We obtained stellar mass data from the Max Planck Institute for Astrophysics - Johns Hopkins University (MPA-JHU) DR10 catalogue \citep{kauffmann_stellar_2003,brinchmann_physical_2004,tremonti_origin_2004} to correlate with the spiral arm numbers in the actual and predicted GZ2 dataset. These data based on the SDSS DR10 are publicly available\footnote{ \url{https://skyserver.sdss.org/dr10/en/help/browser/browser.aspx\#&&history=description+galSpecExtra+U}} along with comprehensive details about the catalogue, as well as the computations and fits of the galaxy physical properties. The values of $M_{*}$ (flagged as \texttt{LGM\_TOT\_P50}) are computed based on theoretical models of stellar populations \citep{kauffmann_stellar_2003}, and assumed a Kroupa Initial Mass Function \citep[Kroupa IMF,][]{kroupa_variation_2001}. To ensure data reliability, we refine the SDSS catalogue by selecting only objects with trustworthy properties, applying the following criteria: \texttt{RELIABLE}$\neq0$, \texttt{Z\_WARNING}$=0$, \texttt{LGM\_TOT\_P50}$\neq-9999$, and $z>0$. The number of images reduced to 11\,291 after cross referencing with the MPA-JHU mass catalogue.

\begin{figure*}[t!]
\includegraphics[width=\linewidth]{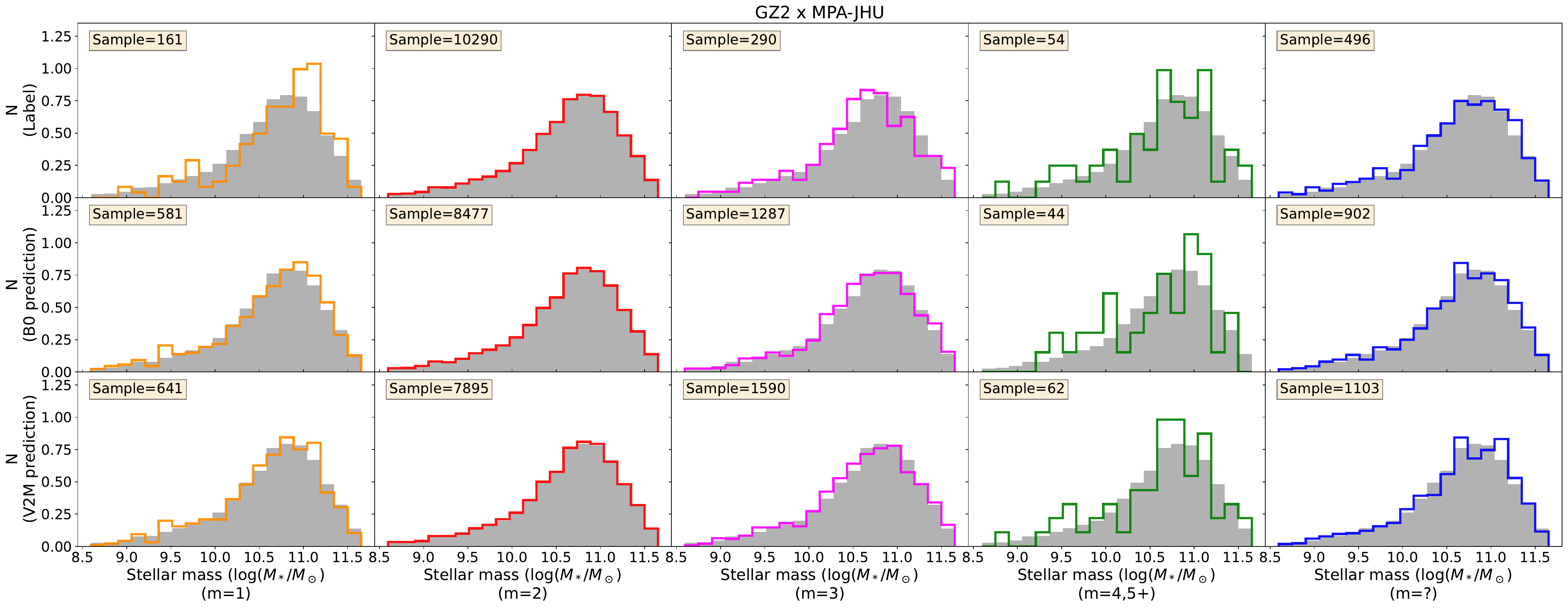}
\caption{Distribution of stellar mass according to spiral arm number obtained from GZ2 labels (top), followed by predictions from the B0 (middle) and V2M (bottom) models. The grey-filled histograms represent the distribution of the entire sample, in comparison with the coloured outlines representing galaxies from each class. It is shown that $m=1$ galaxies have higher stellar masses while $m=3$ galaxies have otherwise.}
\label{reffigure11}
\end{figure*}

\begin{figure*}[t!]
\includegraphics[width=\linewidth]{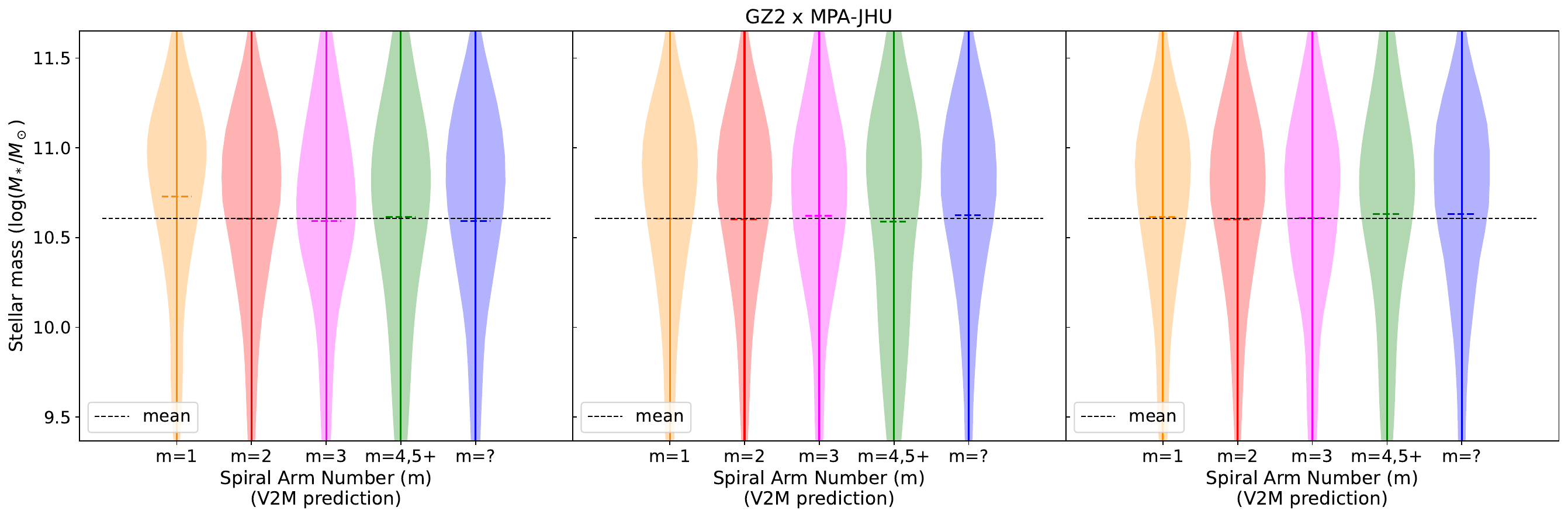}
\caption{Violin plots visualising the normalised stellar mass distribution according to spiral arm number, using GZ2 labels (left), followed by predictions of the B0 (middle) and V2M (right) models. The dotted lines indicate the mean for each distribution. The mean for class $m=1$ in the actual label shifts towards higher stellar mass. Slight shifts above or below overall mean are shown in model predictions for $m=4$,$5+$ and $m=?$.}
\label{reffigure12}
\end{figure*}

The histograms in Figure~\ref{reffigure11} show the distribution of spiral arm numbers for actual labels and predicted classes by the B0 and V2M models, compared to the distribution of the entire dataset. The data distributions are constrained where the sample size must exceed 1\% in the entire dataset for better visualisation. Figure~\ref{reffigure12} illustrates the stellar mass distribution using coloured violin plots, constrained with a threshold of 5\% for the lower and 1\% for the upper data, with the dotted lines representing the mean values.

Distribution difference tests are employed to quantitatively assess the stellar mass tendencies associated with different $m$ classes. The Kolmogorov-Smirnov (K-S) test is a non-parametric statistical test which evaluates the maximum distance between the empirical cumulative distribution functions (CDFs) of the two samples. On the other hand, a one-tailed $t$-test is a statistical test that evaluates the null hypothesis of equal means against the alternative hypothesis that one mean is greater or less than the other. The statistics and $p$-values for both tests are presented in Table~\ref{reftable4}. 

\begin{table*}[t!]
\centering
\begin{tabular}{ccccccc}
\hline
& & & \multicolumn{2}{c}{\textbf{2-Sample K-S Test}} & \multicolumn{2}{c}{\textbf{One-Tailed t-Test}} \\ \cline{4-7} 
& \multirow{-2}{*}{\textbf{m}} & \multirow{-2}{*}{\textbf{Sample size}} & \textbf{Statistic} & \textbf{$p$-value} & \textbf{Statistic} & \textbf{$p$-value} \\ \hline
& 1 & 161 & 0.1196 & \cellcolor[HTML]{D9EAD3}0.0194 & 2.4429 & \cellcolor[HTML]{D9EAD3}0.0146 \\
& 2 & 10290 & 0.0021 & 1.0000 & -0.1116 & 0.9111 \\
& 3 & 290 & 0.0644 & 0.1825 & -0.3531 & 0.7240 \\
& 4,5+ & 54 & 0.0610 & 0.9807 & 0.0946 & 0.9247 \\
\multirow{-5}{*}{\textbf{Label}} & ? & 496 & 0.0262 & 0.9853 & -0.4439 & 0.6571 \\ \hline 
& 1 & 581 & 0.0322 & 0.6209 & -0.0263 & 0.9790 \\
& 2 & 8477 & 0.0036 & 1.0000 & -0.4447 & 0.6565 \\
& 3 & 1287 & 0.0198 & 0.7460 & 0.7917 & 0.4285 \\
& 4,5+ & 44 & 0.1078 & 0.6467 & -0.1826 & 0.8551 \\
\multirow{-5}{*}{\textbf{B0 model}} & ? & 902 & 0.0263 & 0.5990 & 0.8407 & 0.4005 \\ \hline 
& 1 & 641 & 0.0190 & 0.9773 & 0.2750 & 0.7834 \\
& 2 & 7895 & 0.0040 & 1.0000 & -0.4901 & 0.6241 \\
& 3 & 1590 & 0.0218 & 0.5173 & 0.1611 & 0.8721 \\
& 4,5+ & 62 & 0.0628 & 0.9547 & 0.3180 & 0.7505 \\
\multirow{-5}{*}{\textbf{V2M model}} & ? & 1103 & 0.0317 & 0.2597 & 1.1671 & 0.2432 \\ \hline 
\end{tabular}
\caption{Results of the Kolmogorov-Smirnov (K-S) and one-tailed $t$-test for the stellar mass distributions across different spiral arm numbers using the GZ2 labels (top), and predictions from the B0 (middle) and V2M (bottom) models, respectively. Cells shaded in green highlight significant $p$-values which are lower than the threshold of 0.05.}
\label{reftable4}
\end{table*}

The statistics from these tests indicate the degree of similarity between each subsample and parent sample, with smaller values being more similar. The associated $p$-value dictates the significance of the result, with values below the threshold of 0.05 deemed significant. The K-S test compares the cumulative distributions of two dataset to determine the differences (as reflected in Figure~\ref{reffigure11}) while the one-tailed $t$-test compares the mean values (as reflected in Figure~\ref{reffigure12}). A negative $t$-test statistic value indicates a shift towards lower stellar mass, and vice versa.

From Figure~\ref{reffigure11}, we find that there is little evidence for a dependence of spiral arm number with respect to stellar mass. An excess of high-stellar mass galaxies is found for $m=1$, as well as an excess of low-stellar mass galaxies for $m=3$. These are reflected in the K-S statistic in Table~\ref{reftable4}, with $m=1$ having the highest K-S statistic and considered significant due to its $p$-value below 0.05. The class $m=3$ has the second highest K-S statistic but we do not consider it significant due to its relatively higher $p$-value. The violin plots from Figure~\ref{reffigure12} also show a significant tendency towards higher stellar mass in the actual $m=1$ distribution, which is reflected in the high statistic value and $p$-value below 0.05 in the one-tailed $t$-test data. However, $m=3$ does not show a significant difference in the mean value.

The observation that $m=1$ tends to have higher stellar mass and $m=3$ tends to have lower stellar mass (despite its significance level) is consistent with the findings of ref.~\citep{hart_galaxy_2016}. However, it contrasts with the results of ref.~\citep{portertemple_galaxy_2022}, which reported a shift of $m=1$ towards lower stellar mass and $m=3$ towards higher stellar mass. This discrepancy is likely due to the differences of galaxy dataset and stellar mass data used in our study compared to theirs, which will not be further investigated in the current study.

\begin{figure}
\centering
\includegraphics[width=0.7\linewidth]{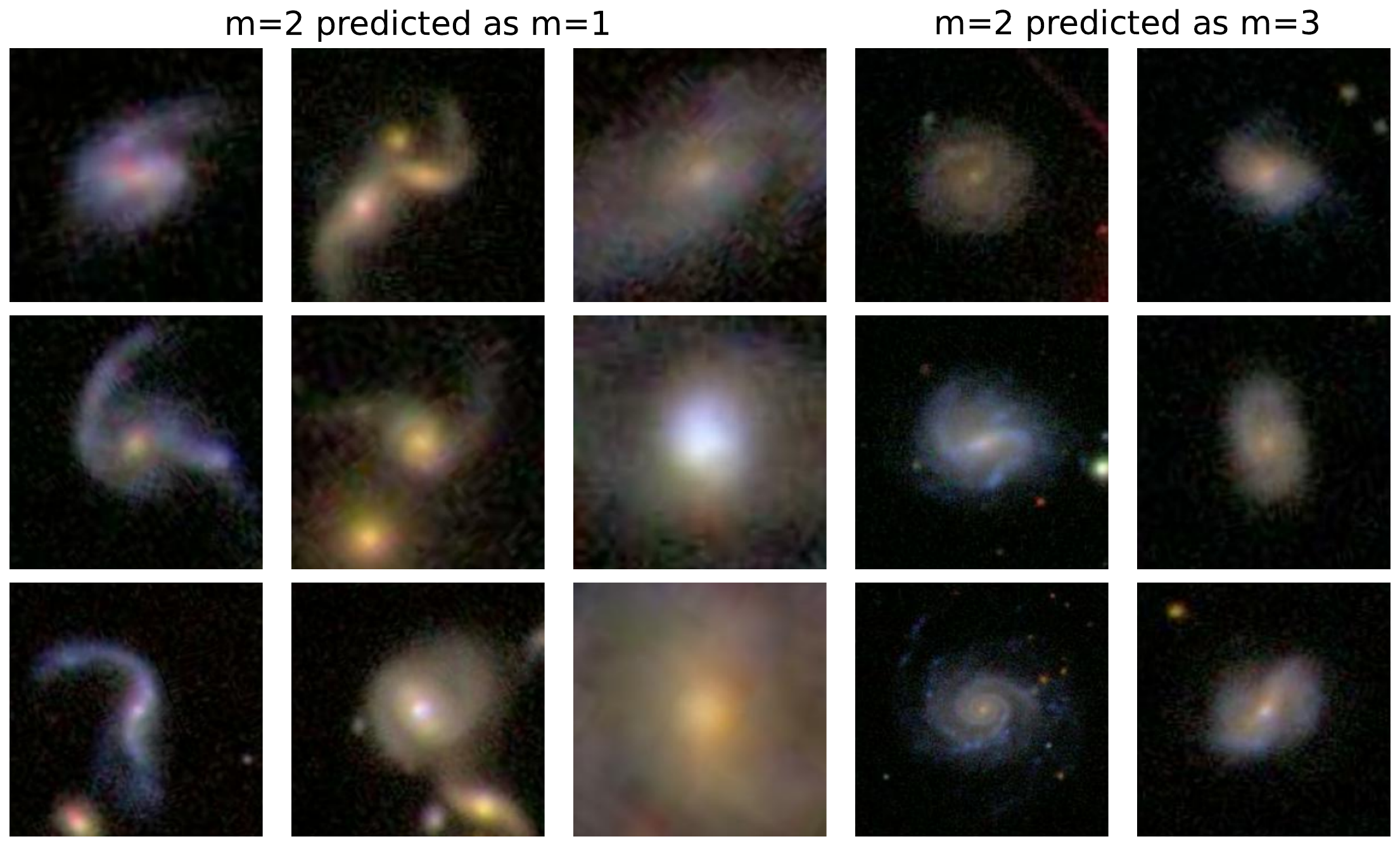}
\caption{Sample images of misclassified $m=2$ galaxies by the V2M model, predicted as $m=1$ (left) and $m=3$ (right). The images consist of asymmetrical spiral structure, external interfering objects, objects being out of sight, indistinguishable spiral arms and diffuse structures.}
\label{reffigure13}
\end{figure}

The stellar mass excesses for $m=1$ and $m=3$ in the histogram are reduced in the model predictions, which is evident from the lower K-S statistic values and higher $p$-values. The violin plots of the model predictions show no significant deviation in the mean stellar mass value for $m=1$ and $m=3$. The reduced tendency disclose the models' misclassification between $m=2$ and $m=1$ or $3$, and the B0 model has fewer misclassifications than the V2M model. Based on Figure~\ref{reffigure13}, the misclassification of $m=2$ galaxies as $m=1$ in the model predictions is primarily due to the asymmetrical spiral structures, external interfering objects, and objects being out of sight. This sheds light on the potential event of tidal interaction. The misclassification of $m=2$ galaxies as $m=3$ is primarily due to the indistinct two-armed or diffuse structures. It is plausible that the $m=2$ and $m=3$ galaxies might be misclassified due the dynamic properties of spiral arms, which may alternate between two and three spiral arms \citep{elmegreen_grand_1993}, and the potential presence of substructure in $m=2$ galaxies \citep{chakrabarti_branch_2003}. Additionally, differences in appearance between optical (showing two arms) and infrared (showing three arms) \citep{block_images_1994} may also contribute to this misclassification.

The mean stellar mass values for $m=?$ in the violin plots shift towards higher mass in both model predictions. However, the $p$-values do not drop below 0.05 despite the high statistic values. The mean stellar mass of $m=4$ and $m=5+$ shifts slightly towards lower mass in the B0 prediction. This differs from the V2M model, where $m=4$ and $m=5+$ shift to higher stellar mass. The histograms also shows a tendency towards higher stellar mass in $m=?$ for the V2M model prediction. However, the $t$-test statistics and $p$-values do not show any significance. Despite its insignificance, the tendency of mean value towards higher stellar mass in high-arm galaxies ($m=4$,$5+$) for the V2M model prediction aligns with the finding of previous studies, which show a trend where more massive galaxies tend to have more spiral arms. This is likely due to the difficulty in detecting high-arm spiral features, which makes high-arm galaxies more noticeable in larger and brighter galaxies.

Overall, the tendency of $m=1$ galaxies towards higher stellar mass and $m=3$ galaxies towards lower stellar mass in the actual GZ2 labels aligns with the results of ref.~\citep{hart_galaxy_2016}. However, these tendencies are diminished in both the B0 and V2M model predictions due to the misclassifications of $m=2$ as $m=1$ and $m=3$. These misclassifications are primarily attributed to the out-of-focus objects, external interference, and indistinct spiral structures. Although the B0 model exhibits lower misclassification rate than the V2M model, the V2M model prediction shows a shift of $m=4$,$5+$ galaxies towards higher stellar mass. This is consistent with the findings of ref.~\citep{hart_galaxy_2016,portertemple_galaxy_2022}, showing a trend where more massive galaxies tend to have more spiral arms. As discussed in Section~\ref{refsec5}, the V2M model demonstrates a higher capability for extracting important features from galaxy images. The model predictions appear to rely predominantly on the image input, as factors such as object interference, field of view and the dynamic spiral arm structure, elements often considered in human decision-making, are not accounted for in CNN model that only take images as input. This observation highlights the potential benefit of incorporating additional parameters in the training process. Despite these limitations, the models, particularly the V2M model, show their effectiveness in classifying galaxy datasets by extracting meaningful features from the image inputs.

\section{Conclusion and Future Work} \label{refsec7}
In this study, we developed CNN-based classifiers using variants of the EfficientNet architecture to categorise spiral galaxies from GZ2 data based on the number of spiral arms. Both models, the V2M fine-tuned on ImageNet and the B0 model using Zoobot pre-trained weights, achieved high classification accuracy, with most performance metrics exceeding 0.8 except for galaxies with 4 spiral arms (class $m=4$). To address challenges in classifying low-population spiral arm classes, we merged higher-arm-number classes ($m=4$, $5+$), which resulted in a 280\% accuracy improvement in the V2M model. This allowed the model to focus on more distinct features across fewer, broader categories with a more balanced class distribution. Although merging the classes $m=3$ and $m=4$ improved the accuracy more significantly, their distinct characteristics make distinguishing between 3 and 4 arms scientifically important. Therefore, while we advocate for the merger of classes to improve the classification results, the choice of merging which classes heavily depend on the purpose of the work involved.

Grad-CAM++ and SmoothGrad were employed to ensure the models were classifying galaxies as expected. Grad-CAM++ showed that the models' decisions were mainly based on the overall structural features of galaxies, while SmoothGrad showed that the models could extract spiral arm features. Lower-arm galaxies are more likely to be predicted as $m=?$ when the spiral arms are less visible, while higher-arm galaxies are more likely to be misclassified as having fewer spiral arm number when the arms are not completely captured. The V2M model has better capabilities in distinguishing the galaxy structure and extracting spiral arms.

The stellar mass distributions of the galaxies were compared across different spiral arm numbers. A significant tendency towards higher stellar mass was observed in $m=1$ galaxies, agreeing with the work of previous research. A tendency towards lower stellar mass was observed in $m=3$ galaxies using the GZ2 label, however this tendency is less obvious in the B0 and V2M model predictions. The weak results from the prediction values primarily lies in the misclassification of $m=2$ galaxies as $m=1$ or $3$, due to their indistinct spiral structures, likely resulting from the external interference ($m=1$) and dynamic nature of spiral arms ($m=3$). A slight tendency towards higher stellar mass in the high-arm galaxies was observed in the V2M model prediction, likely because high-arm galaxies are more noticeable in large, bright galaxies.

The success of CNN in classifying galaxies by spiral arm numbers using a limited dataset, as demonstrated in this study, highlights its strong potential for advancing the morphology classification of spiral galaxies and contributing to the study of spiral arm formation theories. The classifiers introduced in this study are particularly effective for distinguishing low-arm galaxies with distinct spiral structure, and differentiating between low-arm and high-arm galaxies. However, further improvements are needed for accurate categorisation of high-arm galaxies, which requires more dataset for model training. Incorporating additional parameters into the model training could also enhance classification accuracy by accounting for factors such as object interference, field of view and the dynamic spiral structures. Future work should explore the application of these CNN models to larger datasets for assessing their scalability and performance in more extensive and diverse galaxy samples. Additionally, the promising results from using SmoothGrad to extract detailed structural information suggest a valuable avenue for future research, such as differentiating between multiple-arm and flocculent arms.

\acknowledgments

JYHS acknowledges financial support via the Fundamental Research Grant Scheme by the Malaysian Ministry of Higher Education (FRGS/1/2023/STG07/USM/02/14).

\paragraph{Funding Statement} This research was supported by a grant from the Malaysian Ministry of Higher Education (award ID: FRGS/1/2023/STG07/USM/02/14).


\bibliographystyle{JHEP}
\bibliography{biblio.bib}

\providecommand{\href}[2]{#2}\begingroup\raggedright\begin{thebibliography}{10}

\bibitem{hubble_extragalactic_1926}
E.P.~Hubble, \emph{Extragalactic nebulae.}, \href{https://doi.org/10.1086/143018}{\emph{ApJ} {\bfseries 64} (1926) 321}.

\bibitem{kennicutt_star_1998}
R.C.J.~Kennicutt, \emph{Star formation in galaxies along the {Hubble} sequence}, \href{https://doi.org/10.1146/annurev.astro.36.1.189}{\emph{Annual Review of A{\&A}} {\bfseries 36} (1998) 189}.

\bibitem{masters_galaxy_2019}
K.L.~Masters, C.J.~Lintott, R.E.~Hart, S.J.~Kruk, R.J.~Smethurst, K.V.~Casteels et~al., \emph{Galaxy {Zoo}: unwinding the winding problem--observations of spiral bulge prominence and arm pitch angles suggest local spiral galaxies are winding}, \href{https://doi.org/10.1093/mnras/stz1153}{\emph{MNRAS} {\bfseries 487} (2019) 1808}.

\bibitem{devaucouleurs_classification_1959}
G.~{De Vaucouleurs}, \emph{Classification and morphology of external galaxies}, \href{https://doi.org/10.1007/978-3-642-45932-0_7}{\emph{Handbuch der Physik} {\bfseries 53} (1959) 275}.

\bibitem{elmegreen_flocculent_1982}
D.M.~Elmegreen and B.G.~Elmegreen, \emph{Flocculent and grand design spiral structure in field, binary and group galaxies}, \href{https://doi.org/10.1093/mnras/201.4.1021}{\emph{MNRAS} {\bfseries 201} (1982) 1021}.

\bibitem{elmegreen_arm_1987}
D.M.~Elmegreen and B.G.~Elmegreen, \emph{Arm classifications for spiral galaxies}, \href{https://doi.org/10.1086/165034}{\emph{ApJ} {\bfseries 314} (1987) 3}.

\bibitem{lintott_galaxy_2011}
C.~Lintott, K.~Schawinski, S.~Bamford, A.~Slosar, K.~Land, D.~Thomas et~al., \emph{Galaxy {Zoo} 1: data release of morphological classifications for nearly 900000 galaxies}, \href{https://doi.org/10.1111/j.1365-2966.2010.17432.x}{\emph{MNRAS} {\bfseries 410} (2011) 166}.

\bibitem{willett_galaxy_2013}
K.W.~Willett, C.J.~Lintott, S.P.~Bamford, K.L.~Masters, B.D.~Simmons, K.R.V.~Casteels et~al., \emph{Galaxy {Zoo} 2: detailed morphological classifications for 304122 galaxies from the {Sloan} {Digital} {Sky} {Survey}}, \href{https://doi.org/10.1093/mnras/stt1458}{\emph{MNRAS} {\bfseries 435} (2013) 2835}.

\bibitem{lindblad_on_1953}
B.~Lindblad and R.G.~Langebartel, \emph{On the dynamics of stellar systems}, {\emph{Stockholms Observatoriums Annaler} {\bfseries 17} (1953) 6}.

\bibitem{lin_on_1964}
C.C.~Lin and F.H.~Shu, \emph{On the spiral structure of disk galaxies}, \href{https://doi.org/10.1142/9789814415651_0033}{\emph{ApJ} {\bfseries 140} (1964) 646}.

\bibitem{goldreich_spiral_1965}
P.~Goldreich and D.~{Lynden-Bell}, \emph{Ii. spiral arms as sheared gravitational instabilities}, \href{https://doi.org/10.1093/mnras/130.2.125}{\emph{MNRAS} {\bfseries 130} (1965) 125}.

\bibitem{julian_non_1966}
W.H.~Julian and A.~Toomre, \emph{Non-axisymmetric responses of differentially rotating disks of stars}, \href{https://doi.org/10.1086/148957}{\emph{ApJ} {\bfseries 146} (1966) 810}.

\bibitem{toomre_galactic_1972}
A.~Toomre and J.~Toomre, \emph{Galactic bridges and tails}, \href{https://doi.org/10.1086/151823}{\emph{ApJ} {\bfseries 178} (1972) 623}.

\bibitem{hubble_direction_1943}
E.P.~Hubble, \emph{The direction of rotation in spiral nebulae.}, \href{https://doi.org/10.1086/144504}{\emph{ApJ} {\bfseries 97} (1943) 112}.

\bibitem{elmegreen_grand_1993}
B.G.~Elmegreen and M.~Thomasson, \emph{Grand design and flocculent spiral structure in computer simulations with star formation and gas heating}, {\emph{A\&A} {\bfseries 272} (1993) 37}.

\bibitem{zhang_secular_1996}
X.~Zhang, \emph{Secular evolution of spiral galaxies. {I}. a collective dissipation process}, \href{https://doi.org/10.1086/176717}{\emph{ApJ} {\bfseries 457} (1996) 125}.

\bibitem{block_morphological_1991}
D.L.~Block and R.J.~Wainscoatt, \emph{Morphological differences between optical and infrared images of the spiral galaxy {NGC309}}, \href{https://doi.org/10.1038/353048a0}{\emph{Nature} {\bfseries 353} (1991) 48}.

\bibitem{block_images_1994}
D.L.~Block, G.~Bertin, A.~Stockton, P.~Grosb{\o}l, A.F.M.~Moorwood and R.F.~Peletier, \emph{2.1 $\mu$m images of the evolved stellar disk and the morphological classification of spiral galaxies}, {\emph{A\&A} {\bfseries 288} (1994) 365}.

\bibitem{shabani_search_2018}
F.~Shabani, E.K.~Grebel, A.~Pasquali, E.~D{\text{\'O}}nghia, J.S.~Gallagher~{III}, A.~Adamo et~al., \emph{Search for star cluster age gradients across spiral arms of three {LEGUS} disc galaxies}, \href{https://doi.org/10.1093/mnras/sty1277}{\emph{MNRAS} {\bfseries 478} (2018) 3590}.

\bibitem{peterken_direct_2019}
T.G.~Peterken, M.R.~Merrifield, A.~Arag{\'o}n-Salamanca, N.~Drory, C.M.~Krawczyk, K.L.~Masters et~al., \emph{A direct test of density wave theory in a grand-design spiral galaxy}, \href{https://doi.org/10.1038/s41550-018-0627-5}{\emph{Nature Astronomy} {\bfseries 3} (2019) 178}.

\bibitem{bialopetravi_study_2020}
J.~Bialopetravi{\v{c}}ius and D.~Narbutis, \emph{Study of star clusters in the {M83} galaxy with a convolutional neural network}, \href{https://doi.org/10.3847/1538-3881/abbf53}{\emph{AJ} {\bfseries 160} (2020) 264}.

\bibitem{abdeen_evidence_2022}
S.~Abdeen, B.L.~Davis, R.~Eufrasio, D.~Kennefick, J.~Kennefick, R.~Miller et~al., \emph{Evidence in favour of density wave theory through age gradients observed in star formation history maps and spatially resolved stellar clusters}, \href{https://doi.org/10.1093/mnras/stac459}{\emph{MNRAS} {\bfseries 512} (2022) 366}.

\bibitem{kormendy_observational_1979}
J.~Kormendy and C.A.~Norman, \emph{Observational constraints on driving mechanisms for spiral density waves}, \href{https://doi.org/10.1086/157414}{\emph{ApJ} {\bfseries 233} (1979) 539}.

\bibitem{grosbol_spiral_2004}
P.~Grosb{\o}l, P.A.~Patsis and E.~Pompei, \emph{Spiral galaxies observed in the near-infrared {K} band--{I}. {Data} analysis and structural parameters}, \href{https://doi.org/10.1051/0004-6361:20035804}{\emph{A\&A} {\bfseries 423} (2004) 849}.

\bibitem{grosbol_galactic_2018}
P.~Grosb{\o}l and G.~Carraro, \emph{Is the {Galactic} {Spiral} {Potential} {2}-or {4}-arms?}, \href{https://doi.org/10.1017/S174392131700650}{\emph{Proc. {IAU}} {\bfseries 334} (2018) 300}.

\bibitem{khrapov_modeling_2021}
S.~Khrapov, A.~Khoperskov and V.~Korchagin, \emph{Modeling of spiral structure in a multi-component {Milky} {Way}-like galaxy}, \href{https://doi.org/10.3390/galaxies9020029}{\emph{Galaxies} {\bfseries 9} (2021) 29}.

\bibitem{hart_galaxy_2018}
R.E.~Hart, S.P.~Bamford, W.C.~Keel, S.J.~Kruk, K.L.~Masters, B.D.~Simmons et~al., \emph{Galaxy {Zoo}: constraining the origin of spiral arms}, \href{https://doi.org/10.1093/mnras/sty1201}{\emph{MNRAS} {\bfseries 478} (2018) 932}.

\bibitem{hart_galaxy_2016}
R.E.~Hart, S.P.~Bamford, K.W.~Willett, K.L.~Masters, C.~Cardamone, C.J.~Lintott et~al., \emph{Galaxy {Zoo}: comparing the demographics of spiral arm number and a new method for correcting redshift bias}, \href{https://doi.org/10.1093/mnras/stw1588}{\emph{MNRAS} {\bfseries 461} (2016) 3663}.

\bibitem{vandenbergh_new_1976}
S.~{van den Bergh}, \emph{A new classification system for galaxies}, \href{https://doi.org/10.1086/154452}{\emph{ApJ} {\bfseries 206} (1976) 883}.

\bibitem{abraham_morphologies_1994}
R.G.~Abraham, F.~Valdes, H.K.C.~Yee and S.~{van den Bergh}, \emph{The morphologies of distant galaxies. {I}. an automated classification system}, \href{https://doi.org/10.1086/174550}{\emph{ApJ} {\bfseries 432} (1994) 75}.

\bibitem{abraham_galaxy_1996}
R.G.~Abraham, N.R.~Tanvir, B.X.~Santiago, R.S.~Ellis, K.~Glazebrook and S.~{van den Bergh}, \emph{Galaxy morphology to {I}=25 mag in the {Hubble} {Deep} {Field}}, \href{https://doi.org/10.1093/mnras/279.3.L47}{\emph{MNRAS} {\bfseries 279} (1996) L47}.

\bibitem{abraham_morphologies_1996}
R.G.~{Abraham}, S.~{van den Bergh}, K.~{Glazebrook}, R.S.~{Ellis}, B.X.~{Santiago}, P.~{Surma} et~al., \emph{The morphologies of distant galaxies. {II}. {Classifications} from the {Hubble} {Space} {Telescope} {Medium} {Deep} {Survey}}, \href{https://doi.org/10.1086/192352}{\emph{ApJSS} {\bfseries 107} (1996) 1}.

\bibitem{conselice_asymmetry_2000}
C.J.~{Conselice}, M.A.~{Bershady} and A.~{Jangren}, \emph{The asymmetry of galaxies: Physical morphology for nearby and high-redshift galaxies}, \href{https://doi.org/10.1086/308300}{\emph{ApJ} {\bfseries 529} (2000) 886}.

\bibitem{naim_automated_1995}
A.~{Naim}, O.~{Lahav}, L.J.~{Sodre} and M.C.~{Storrie-Lombardi}, \emph{Automated morphological classification of {APM} galaxies by supervised artificial neural networks}, \href{https://doi.org/10.1093/mnras/275.3.567}{\emph{MNRAS} {\bfseries 275} (1995) 567}.

\bibitem{lahav_neural_1996}
O.~{Lahav}, A.~{Naim}, L.J.~{Sodr{\'e}} and M.C.~{Storrie-Lombardi}, \emph{Neural computation as a tool for galaxy classification: methods and examples}, \href{https://doi.org/10.1093/mnras/283.1.207}{\emph{MNRAS} {\bfseries 283} (1996) 207}.

\bibitem{owens_oblique_1996}
E.A.~{Owens}, R.E.~{Griffiths} and K.U.~{Ratnatunga}, \emph{Using oblique decision trees for the morphological classification of galaxies}, \href{https://doi.org/10.1093/mnras/281.1.153}{\emph{MNRAS} {\bfseries 281} (1996) 153}.

\bibitem{bazell_ensembles_2001}
D.~{Bazell} and D.W.~{Aha}, \emph{Ensembles of classifiers for morphological galaxy classification}, \href{https://doi.org/10.1086/318696}{\emph{ApJ} {\bfseries 548} (2001) 219}.

\bibitem{goderya_morphological_2002}
S.N.~{Goderya} and S.M.~{Lolling}, \emph{Morphological classification of galaxies using computer vision and artificial neural networks: A computational scheme}, \href{https://doi.org/10.1023/A:1015193432240}{\emph{A\&SS} {\bfseries 279} (2002) 377}.

\bibitem{ball_galaxy_2004}
N.M.~{Ball}, J.~{Loveday}, M.~{Fukugita}, O.~{Nakamura}, S.~{Okamura}, J.~{Brinkmann} et~al., \emph{Galaxy types in the {Sloan} {Digital} {Sky} {Survey} using supervised artificial neural networks}, \href{https://doi.org/10.1111/j.1365-2966.2004.07429.x}{\emph{MNRAS} {\bfseries 348} (2004) 1038}.

\bibitem{huertascompany_robust_2008}
M.~{Huertas-Company}, D.~{Rouan}, L.~{Tasca}, G.~{Soucail} and O.~{Le F{\`e}vre}, \emph{A robust morphological classification of high-redshift galaxies using support vector machines on seeing limited images. {I}. method description}, \href{https://doi.org/10.1051/0004-6361:20078625}{\emph{A\&A} {\bfseries 478} (2008) 971}.

\bibitem{huertascompany_robust_2009}
M.~{Huertas-Company}, L.~{Tasca}, D.~{Rouan}, D.~{Pelat}, J.P.~{Kneib}, O.~{Le F{\`e}vre} et~al., \emph{A robust morphological classification of high-redshift galaxies using support vector machines on seeing limited images. {II}. {Quantifying} morphological k-correction in the {COSMOS} field at 1 {\textless} z {\textless} 2: {Ks} band vs. {I} band}, \href{https://doi.org/10.1051/0004-6361/200811255}{\emph{A\&A} {\bfseries 497} (2009) 743}.

\bibitem{huertascompany_revisitng_2011}
M.~{Huertas-Company}, J.A.L.~{Aguerri}, M.~{Bernardi}, S.~{Mei} and J.~{S{\'a}nchez Almeida}, \emph{Revisiting the {Hubble} sequence in the {SDSS} {DR7} spectroscopic sample: a publicly available {Bayesian} automated classification}, \href{https://doi.org/10.1051/0004-6361/201015735}{\emph{A\&A} {\bfseries 525} (2011) A157}.

\bibitem{banerji_galaxy_2010}
M.~{Banerji}, O.~{Lahav}, C.J.~{Lintott}, F.B.~{Abdalla}, K.~{Schawinski}, S.P.~{Bamford} et~al., \emph{Galaxy {Zoo}: reproducing galaxy morphologies via machine learning}, \href{https://doi.org/10.1111/j.1365-2966.2010.16713.x}{\emph{MNRAS} {\bfseries 406} (2010) 342}.

\bibitem{york_sloan_2000}
D.G.~York, J.~Adelman, J.E.~Anderson~Jr, S.F.~Anderson, J.~Annis, N.A.~Bahcall et~al., \emph{The sloan digital sky survey: {Technical} summary}, \href{https://doi.org/10.1086/301513}{\emph{AJ} {\bfseries 120} (2000) 1579}.

\bibitem{cheng_optimizing_2020}
T.-Y.~Cheng, C.J.~Conselice, A.~{Arag{\'o}n-Salamanca}, N.~Li, A.F.L.~Bluck, W.G.~Hartley et~al., \emph{Optimizing automatic morphological classification of galaxies with machine learning and deep learning using {Dark} {Energy} {Survey} imaging}, \href{https://doi.org/10.1093/mnras/staa501}{\emph{MNRAS} {\bfseries 493} (2020) 4209}.

\bibitem{dieleman_rotation_2015}
S.~{Dieleman}, K.W.~{Willett} and J.~{Dambre}, \emph{Rotation-invariant convolutional neural networks for galaxy morphology prediction}, \href{https://doi.org/10.1093/mnras/stv632}{\emph{MNRAS} {\bfseries 450} (2015) 1441}.

\bibitem{dominguezsanchez_improving_2018}
H.~{Dom{\'i}nguez S{\'a}nchez}, M.~{Huertas-Company}, M.~{Bernardi}, D.~{Tuccillo} and J.L.~{Fischer}, \emph{Improving galaxy morphologies for {SDSS} with deep learning}, \href{https://doi.org/10.1093/mnras/sty338}{\emph{MNRAS} {\bfseries 476} (2018) 3661}.

\bibitem{zhu_galaxy_2019}
X.-P.~{Zhu}, J.-M.~{Dai}, C.-J.~{Bian}, Y.~{Chen}, S.~{Chen} and C.~{Hu}, \emph{Galaxy morphology classification with deep convolutional neural networks}, \href{https://doi.org/10.1007/s10509-019-3540-1}{\emph{A\&SS} {\bfseries 364} (2019) 55}.

\bibitem{kalvankar_galaxy_2020}
S.~Kalvankar, H.~Pandit and P.~Parwate, \emph{Galaxy morphology classification using {EfficientNet} architectures}, {\emph{ArXiv e-prints} (2020) } [\href{https://arxiv.org/abs/2008.13611}{{\ttfamily 2008.13611}}].

\bibitem{cavanagh_morphological_2021}
M.K.~Cavanagh, K.~Bekki and B.A.~Groves, \emph{Morphological classification of galaxies with deep learning: comparing 3-way and 4-way {CNNs}}, \href{https://doi.org/10.1093/mnras/stab1552}{\emph{MNRAS} {\bfseries 506} (2021) 659}.

\bibitem{hart_galaxy_2017}
R.E.~Hart, S.P.~Bamford, W.B.~Hayes, C.N.~Cardamone, W.C.~Keel, S.J.~Kruk et~al., \emph{Galaxy {Zoo} and {SPARCFIRE}: constraints on spiral arm formation mechanisms from spiral arm number and pitch angles}, \href{https://doi.org/10.1093/mnras/stx2137}{\emph{MNRAS} {\bfseries 472} (2017) 2263}.

\bibitem{davis_sparcfire_2014}
D.R.~Davis and W.B.~Hayes, \emph{{SpArcFiRe}: scalable automated detection of spiral galaxy arm segments}, \href{https://doi.org/10.1088/0004-637X/790/2/87}{\emph{ApJ} {\bfseries 790} (2014) 87}.

\bibitem{bekki_quantifying_2021}
K.~Bekki, \emph{Quantifying the fine structures of disk galaxies with deep learning: {Segmentation} of spiral arms in different {Hubble} types}, \href{https://doi.org/10.1051/0004-6361/202039797}{\emph{A\&A} {\bfseries 647} (2021) A120}.

\bibitem{ronneberger_unet_2015}
O.~Ronneberger, P.~Fischer and T.~Brox, \emph{{U-net}: Convolutional networks for biomedical image segmentation},  in \emph{Medical image computing and computer-assisted intervention--{MICCAI} 2015}, pp.~234--241, 2015, \href{https://doi.org/10.1007/978-3-319-24574-4_28}{DOI}.

\bibitem{laureijs_euclid_2011}
R.~Laureijs, J.~Amiaux, S.~Arduini, J.-L.~Augueres, J.~Brinchmann, R.~Cole et~al., \emph{Euclid definition study report}, {\emph{ArXiv e-prints} (2011) } [\href{https://arxiv.org/abs/1110.3193}{{\ttfamily 1110.3193}}].

\bibitem{spergel_wide_2015}
D.~Spergel, N.~Gehrels, C.~Baltay, D.~Bennett, J.~Breckinridge, M.~Donahue et~al., \emph{Wide-field infrarred survey telescope-astrophysics focused telescope assets {WFIRST-AFTA} 2015 report}, {\emph{ArXiv e-prints} (2015) } [\href{https://arxiv.org/abs/1503.03757}{{\ttfamily 1503.03757}}].

\bibitem{ivezic_lsst_2019}
{\v{Z}}.~{Ivezi\'c}, S.M.~Kahn, J.A.~Tyson, B.~Abel, E.~Acosta, R.~Allsman et~al., \emph{{LSST}: from science drivers to reference design and anticipated data products}, \href{https://doi.org/10.3847/1538-4357/ab042c}{\emph{ApJ} {\bfseries 873} (2019) 111}.

\bibitem{lintott_galaxy_2008}
C.J.~Lintott, K.~Schawinski, A.~Slosar, K.~Land, S.~Bamford, D.~Thomas et~al., \emph{Galaxy {Zoo}: morphologies derived from visual inspection of galaxies from the {Sloan} {Digital} {Sky} {Survey}}, \href{https://doi.org/10.1111/j.1365-2966.2008.13689.x}{\emph{MNRAS} {\bfseries 389} (2008) 1179}.

\bibitem{annis_sloan_2014}
J.~Annis, M.~Soares-Santos, M.A.~Strauss, A.C.~Becker, S.~Dodelson, X.~Fan et~al., \emph{The sloan digital sky survey coadd: {275} {deg2} of deep sloan digital sky survey imaging on stripe {82}}, \href{https://doi.org/10.1088/0004-637X/794/2/120}{\emph{ApJ} {\bfseries 794} (2014) 120}.

\bibitem{lecun_handwritten_1989}
Y.~{Le Cun}, L.D.~Jackel, B.~Boser, J.S.~Denker, H.P.~Graf, I.~Guyon et~al., \emph{Handwritten digit recognition: applications of neural network chips and automatic learning}, \href{https://doi.org/10.1109/35.41400}{\emph{{IEEE} {Communications} {Magazine}} {\bfseries 27} (1989) 41}.

\bibitem{goodfellow_deep_2016}
I.~Goodfellow, Y.~Bengio and A.~Courville, \emph{Deep learning}, {MIT} press (2016).

\bibitem{lecun_gradient_1998}
Y.~LeCun, L.~Bottou, Y.~Bengio and P.~Haffner, \emph{Gradient-based learning applied to document recognition}, \href{https://doi.org/10.1109/5.726791}{\emph{Proc. {IEEE}} {\bfseries 86} (1998) 2278}.

\bibitem{gu_recent_2018}
J.~Gu, Z.~Wang, J.~Kuen, L.~Ma, A.~Shahroudy, B.~Shuai et~al., \emph{Recent advances in convolutional neural networks}, \href{https://doi.org/10.1016/j.patcog.2017.10.013}{\emph{Pattern {Recognition}} {\bfseries 77} (2018) 354}.

\bibitem{he_deep_2016}
K.~He, X.~Zhang, S.~Ren and J.~Sun, \emph{Deep residual learning for image recognition},  in \emph{2016 {IEEE} {Conference} on {Computer} {Vision} and {Pattern} {Recognition}}, pp.~770--778, 2016, \href{https://doi.org/10.1109/CVPR.2016.90}{DOI}.

\bibitem{tan_efficientnet_2019}
M.~Tan and Q.~Le, \emph{{EfficientNet}: Rethinking model scaling for convolutional neural networks},  in \emph{Proceedings of the 36th {International} {Conference} on {Machine} {Learning}}, vol.~97, pp.~6105--6114, 2019.

\bibitem{sandler_mobilenetv2_2018}
M.~Sandler, A.~Howard, M.~Zhu, A.~Zhmoginov and L.-C.~Chen, \emph{{MobileNetV2}: Inverted residuals and linear bottlenecks},  in \emph{2018 {IEEE/CVF} {Conference} on {Computer} {Vision} and {Pattern} {Recognition}}, pp.~4510--4520, 2018, \href{https://doi.org/10.1109/CVPR.2018.00474}{DOI}.

\bibitem{hu_squeeze_2018}
J.~Hu, L.~Shen and G.~Sun, \emph{Squeeze-and-excitation networks},  in \emph{2018 {IEEE/CVF} {Conference} on {Computer} {Vision} and {Pattern} {Recognition}}, pp.~7132--7141, 2018, \href{https://doi.org/10.1109/CVPR.2018.00745}{DOI}.

\bibitem{tan_efficientnetv2_2021}
M.~Tan and Q.~Le, \emph{{EfficientNetV2}: Smaller models and faster training},  in \emph{Proceedings of the 38th {International} {Conference} on {Machine} {Learning}}, vol.~139, pp.~10096--10106, 2021.

\bibitem{deng_imagenet_2009}
J.~Deng, W.~Dong, R.~Socher, L.-J.~Li, K.~Li and F.-F.~Li, \emph{{ImageNet}: A large-scale hierarchical image database},  in \emph{2009 {IEEE} {Conference} on {Computer} {Vision} and {Pattern} {Recognition}}, pp.~248--255, 2009, \href{https://doi.org/10.1109/CVPR.2009.5206848}{DOI}.

\bibitem{walmsley_zoobot_2023}
M.~Walmsley, C.~Allen, B.~Aussel, M.~Bowles, K.~Gregorowicz, I.V.~Slijepcevic et~al., \emph{{Zoobot}: Adaptable deep learning models for galaxy morphology}, \href{https://doi.org/10.21105/joss.05312}{\emph{Journal of Open Source Software} {\bfseries 8} (2023) 5312}.

\bibitem{huang_densely_2017}
G.~Huang, Z.~Liu, L.~{van der Maaten} and K.Q.~Weinberger, \emph{Densely connected convolutional networks},  in \emph{2017 {IEEE} {Conference} on {Computer} {Vision} and {Pattern} {Recognition}}, pp.~4700--4708, 2017, \href{https://doi.org/10.1109/CVPR.2017.243}{DOI}.

\bibitem{walmsley_galaxy_2022}
M.~Walmsley, C.~Lintott, T.~G\'eron, S.~Kruk, C.~Krawczyk, K.W.~Willett et~al., \emph{Galaxy {Zoo} {DECaLS}: Detailed visual morphology measurements from volunteers and deep learning for {314 000} galaxies}, \href{https://doi.org/10.1093/mnras/stab2093}{\emph{MNRAS} {\bfseries 509} (2022) 3966}.

\bibitem{walmsley_practical_2022}
M.~Walmsley, A.M.M.~Scaife, C.~Lintott, M.~Lochner, V.~Etsebeth, T.~G\'eron et~al., \emph{Practical galaxy morphology tools from deep supervised representation learning}, \href{https://doi.org/10.1093/mnras/stac525}{\emph{MNRAS} {\bfseries 513} (2022) 1581}.

\bibitem{etsebeth_astronomaly_2024}
V.~Etsebeth, M.~Lochner, M.~Walmsley and M.~Grespan, \emph{Astronomaly at scale: searching for anomalies amongst 4 million galaxies}, \href{https://doi.org/10.1093/mnras/stae496}{\emph{MNRAS} {\bfseries 529} (2024) 732}.

\bibitem{popp_transfer_2024}
J.J.~Popp, H.~Dickinson, S.~Serjeant, M.~Walmsley, D.~Adams, L.~Fortson et~al., \emph{Transfer learning for galaxy feature detection: {Finding} giant star-forming clumps in low-redshift galaxies using faster region-based convolutional neural network}, \href{https://doi.org/10.1093/rasti/rzae013}{\emph{RASTI} {\bfseries 3} (2024) 174}.

\bibitem{agarap_deep_2018}
A.F.~Agarap, \emph{Deep learning using rectified linear units ({ReLU})}, {\emph{ArXiv e-prints} (2018) } [\href{https://arxiv.org/abs/1803.08375}{{\ttfamily 1803.08375}}].

\bibitem{bridle_training_1989}
J.~Bridle, \emph{Training stochastic model recognition algorithms as networks can lead to maximum mutual information estimation of parameters},  in \emph{{Advances} in {Neural} {Information} {Processing} {Systems}}, vol.~2, 1989.

\bibitem{chollet_keras_2015}
F.~Chollet et~al., \emph{Keras},  2015.

\bibitem{abadi_tensorflow_2016}
M.~Abadi, A.~Agarwal, P.~Barham, E.~Brevdo, Z.~Chen, C.~Citro et~al., \emph{{Tensorflow}: {Large}-scale machine learning on heterogeneous distributed systems}, {\emph{ArXiv e-prints} (2016) } [\href{https://arxiv.org/abs/1603.04467}{{\ttfamily 1603.04467}}].

\bibitem{nvidia_cuda_2020}
{NVIDIA}, P.~Vingelmann and F.H.P.~Fitzek, \emph{{CUDA}, release: 10.2.89},  2020.

\bibitem{kingma_adam_2017}
D.P.~Kingma and J.~Ba, \emph{Adam: {A} method for stochastic optimization}, {\emph{ArXiv e-prints} (2017) } [\href{https://arxiv.org/abs/1412.6980}{{\ttfamily 1412.6980}}].

\bibitem{chattopadhay_gradcam_2018}
A.~Chattopadhay, A.~Sarkar, P.~Howlader and V.N.~Balasubramanian, \emph{{Grad-CAM++}: Generalized gradient-based visual explanations for deep convolutional networks},  in \emph{2018 {IEEE} {Winter} {Conference} on {Applications} of {Computer} {Vision}}, pp.~839--847, 2018, \href{https://doi.org/10.1109/WACV.2018.00097}{DOI}.

\bibitem{smilkov_smoothgrad_2017}
D.~Smilkov, N.~Thorat, B.~Kim, F.~Vi{\'e}gas and M.~Wattenberg, \emph{{SmoothGrad}: removing noise by adding noise}, {\emph{ArXiv e-prints} (2017) } [\href{https://arxiv.org/abs/1706.03825}{{\ttfamily 1706.03825}}].

\bibitem{gordon_uncovering_2024}
A.J.~Gordon, A.~Ferguson and R.G.~Mann, \emph{Uncovering tidal treasures: Automated classification of faint tidal features in {DECaLS} data}, \href{https://doi.org/10.1093/mnras/stae2169}{\emph{MNRAS} {\bfseries 534} (2024) 1459}.

\bibitem{medinarosales_mitigating_2024}
E.~{Medina-Rosales}, G.~{Cabrera-Vives} and C.J.~Miller, \emph{Mitigating bias in deep learning: training unbiased models on biased data for the morphological classification of galaxies}, \href{https://doi.org/10.1093/mnras/stae1088}{\emph{MNRAS} {\bfseries 531} (2024) 52}.

\bibitem{bhambra_explaining_2022}
P.~Bhambra, B.~Joachimi and O.~Lahav, \emph{Explaining deep learning of galaxy morphology with saliency mapping}, \href{https://doi.org/10.1093/mnras/stac368}{\emph{MNRAS} {\bfseries 511} (2022) 5032}.

\bibitem{bamford_galaxy_2009}
S.P.~Bamford, R.C.~Nichol, I.K.~Baldry, K.~Land, C.J.~Lintott, K.~Schawinski et~al., \emph{Galaxy {Zoo}: the dependence of morphology and colour on environment}, \href{https://doi.org/10.1111/j.1365-2966.2008.14252.x}{\emph{MNRAS} {\bfseries 393} (2009) 1324}.

\bibitem{kelvin_galaxy_2014}
L.S.~Kelvin, S.P.~Driver, A.S.G.~Robotham, E.N.~Taylor, A.W.~Graham, M.~Alpaslan et~al., \emph{{Galaxy} {And} {Mass} {Assembly} ({GAMA}): stellar mass functions by {Hubble} type}, \href{https://doi.org/10.1093/mnras/stu1507}{\emph{MNRAS} {\bfseries 444} (2014) 1647}.

\bibitem{munozmateos_spitzer_2015}
J.C.~{Mu{\~n}oz-Mateos}, K.~Sheth, M.~Regan, T.~Kim, J.~Laine, S.~{Erroz-Ferrer} et~al., \emph{The {Spitzer} {Survey} of stellar structure in galaxies ({S4G}): stellar masses, sizes, and radial profiles for 2352 nearby galaxies}, \href{https://doi.org/10.1088/0067-0049/219/1/3}{\emph{ApJSS} {\bfseries 219} (2015) 3}.

\bibitem{kendall_spiral_2015}
S.~Kendall, C.~Clarke and R.C.J.~Kennicutt, \emph{Spiral structure in nearby galaxies -- {II}. {Comparative} analysis and conclusions}, \href{https://doi.org/10.1093/mnras/stu2431}{\emph{MNRAS} {\bfseries 446} (2015) 4155}.

\bibitem{chang_stellar_2015}
Y.-Y.~Chang, A.~{van der Wel}, E.~{da Cunha} and H.-W.~Rix, \emph{Stellar masses and star formation rates for {1 M} galaxies from {SDSS+WISE}}, \href{https://doi.org/10.1088/0067-0049/219/1/8}{\emph{ApJSS} {\bfseries 219} (2015) 8}.

\bibitem{portertemple_galaxy_2022}
R.~{Porter-Temple}, B.W.~Holwerda, A.M.~Hopkins, L.E.~Porter, C.~Henry, T.~Geron et~al., \emph{Galaxy {And} {Mass} {Assembly}: {Galaxy} {Zoo} spiral arms and star formation rates}, \href{https://doi.org/10.1093/mnras/stac1936}{\emph{MNRAS} {\bfseries 515} (2022) 3875}.

\bibitem{dejong_third_2017}
J.T.A.~{de Jong}, G.A.V.~Kleijn, T.~Erben, H.~Hildebrandt, K.~Kuijken, G.~Sikkema et~al., \emph{The third data release of the {Kilo-Degree} {Survey} and associated data products}, \href{https://doi.org/10.1051/0004-6361/201730747}{\emph{A\&A} {\bfseries 604} (2017) A134}.

\bibitem{dacunha_simple_2008}
E.~Da~Cunha, S.~Charlot and D.~Elbaz, \emph{A simple model to interpret the ultraviolet, optical and infrared emission from galaxies}, \href{https://doi.org/10.1111/j.1365-2966.2008.13535.x}{\emph{MNRAS} {\bfseries 388} (2008) 1595}.

\bibitem{kauffmann_stellar_2003}
G.~Kauffmann, T.M.~Heckman, S.D.M.~White, S.~Charlot, C.~Tremonti, J.~Brinchmann et~al., \emph{Stellar masses and star formation histories for 105 galaxies from the {Sloan} {Digital} {Sky} {Survey}}, \href{https://doi.org/10.1046/j.1365-8711.2003.06291.x}{\emph{MNRAS} {\bfseries 341} (2003) 33}.

\bibitem{brinchmann_physical_2004}
J.~Brinchmann, S.~Charlot, S.D.M.~White, C.~Tremonti, G.~Kauffmann, T.~Heckman et~al., \emph{The physical properties of star-forming galaxies in the low-redshift universe}, \href{https://doi.org/10.1111/j.1365-2966.2004.07881.x}{\emph{MNRAS} {\bfseries 351} (2004) 1151}.

\bibitem{tremonti_origin_2004}
C.A.~Tremonti, T.M.~Heckman, G.~Kauffmann, J.~Brinchmann, S.~Charlot, S.D.M.~White et~al., \emph{The origin of the mass-metallicity relation: insights from {53,000} star-forming galaxies in the {Sloan} {Digital} {Sky} {Survey}}, \href{https://doi.org/10.1086/423264}{\emph{ApJ} {\bfseries 613} (2004) 898}.

\bibitem{kroupa_variation_2001}
P.~Kroupa, \emph{On the variation of the initial mass function}, \href{https://doi.org/10.1046/j.1365-8711.2001.04022.x}{\emph{MNRAS} {\bfseries 322} (2001) 231}.

\bibitem{chakrabarti_branch_2003}
S.~Chakrabarti, G.~Laughlin and F.H.~Shu, \emph{Branch, spur, and feather formation in spiral galaxies}, \href{https://doi.org/10.1086/377578}{\emph{ApJ} {\bfseries 596} (2003) 220}.

\end{thebibliography}\endgroup


\end{document}